\documentclass[
  twocolumn,
  prb,
  showpacs,
  amsmath,
  amssymb,
  superscriptaddress,
  floatfix
]{revtex4}

\usepackage{bm}
\usepackage{graphicx}
\usepackage{bbold}

\begin{document}

\title{Ballistic transport in disordered graphene}

\author{A.~Schuessler}
\affiliation{
 Institut f\"ur Nanotechnologie, Forschungszentrum Karlsruhe,
 76021 Karlsruhe, Germany
}

\author{P.~M.~Ostrovsky}
\affiliation{
 Institut f\"ur Nanotechnologie, Forschungszentrum Karlsruhe,
 76021 Karlsruhe, Germany
}
\affiliation{
 L.~D.~Landau Institute for Theoretical Physics RAS,
 119334 Moscow, Russia
}

\author{I.~V.~Gornyi}
\affiliation{
 Institut f\"ur Nanotechnologie, Forschungszentrum Karlsruhe,
 76021 Karlsruhe, Germany
}
\affiliation{
 A.F.~Ioffe Physico-Technical Institute,
 194021 St.~Petersburg, Russia.
}

\author{A.~D.~Mirlin}
\affiliation{
 Institut f\"ur Nanotechnologie, Forschungszentrum Karlsruhe,
 76021 Karlsruhe, Germany
}
\affiliation{
 Inst. f\"ur Theorie der kondensierten Materie,
 Universit\"at Karlsruhe, 76128 Karlsruhe, Germany
}
\affiliation{
 Petersburg Nuclear Physics Institute,
 188300 St.~Petersburg, Russia.
}

\begin{abstract}
An analytic theory of electron transport in disordered graphene in a ballistic
geometry is developed. We consider a sample of a large width $W$ and analyze
the evolution of the conductance, the shot noise, and the full statistics of the
charge transfer with increasing length $L$, both at the Dirac point and at a
finite gate voltage. The transfer matrix approach combined with the disorder
perturbation theory and the renormalization group is used. We also discuss the
crossover to the diffusive regime and construct a ``phase diagram'' of various
transport regimes in graphene.
\end{abstract}


\maketitle

\section{Introduction}

Recent successes in manufacturing of atomically thin graphite samples
\cite{Novoselov04, NovoselovPNAS, Novoselov05, Zhang05} (graphene) have
stimulated intense experimental and theoretical activity \cite{Geim07, RMP07}.
The key feature of graphene is the massless Dirac type of low-energy electron
excitations. This gives rise to a number of remarkable physical properties of
this system distinguishing it from conventional two-dimensional metals. One of
the most prominent features of graphene is the ``minimal conductivity'' at the
neutrality (Dirac) point. Specifically, the conductivity\cite{Novoselov05,
Zhang05,Kim} of an undoped sample is close to $e^2/h$ per spin per valley,
remaining almost constant in a very broad temperature range --- from room
temperature down to 30mK.

Several recent theoretical works addressed transport in disordered graphene
samples. It was found that localization properties depend strongly on the nature
of disorder \cite{Aleiner06, McCann, Altland06, OurPRB, OurPRL, OurEPJ, OurQHE}
which determines the symmetry and topology of the corresponding field theory.
The localization is absent provided a certain symmetry of clean graphene
Hamiltonian is preserved in the disordered sample, see Ref.\
\onlinecite{OurEPJ}.

One possibility is that disorder preserves a chiral symmetry of massless Dirac
fermions. This situation is, in particular, realized when the dominant disorder
is due to corrugations of graphene sheet (ripples) and/or dislocations.
\cite{GuineaMorpurgo} The conductivity in  such chiral-symmetric models has been
shown \cite{OurPRB} to be exactly $e^2/\pi h$ (per spin per valley) at the Dirac
point. While being temperature independent, this value is, however, less by a
factor of $\sim 3$ than the experimentally measured values.

Another possibility, a long-range randomness, was studied in Ref.\
\onlinecite{OurPRL}. This type of disorder does not mix the two valleys of the
graphene spectrum, which leads to emergence of a topological term in the
corresponding field theory (unitary or symplectic $\sigma$-model). The peculiar
topological properties protect the system from localization \cite{OurPRL,
OurEPJ, OurQHE, Ryu07, Koshino07}. It is worth mentioning that a topologically
protected metallic state emerging in graphene with long-range random potential
also arises at a surface of a three-dimensional $Z_2$ topological insulator.
\cite{Hasan08, Schnyder08}

A number of numerical simulations of electron transport in disordered graphene
\cite{Nomura07, Rycerz07, Bardarson07, Koshino07, San-Jose07, Lewenkopf08}
confirmed the absence of localization in the presence of long-range random
potential. The main quantity studied numerically in most of these works is the
conductance $G$ of a finite-size graphene sample with a width $W$ much larger
than the length $L$. This setup allows one to define the ``conductivity''
$\sigma\equiv G L/W$ even for  ballistic samples with $L$ much shorter than the
mean free path $l$. Remarkably, in graphene at the Dirac point, such ballistic
``conductivity'' has a universal value $e^2/\pi h$ in the clean case.
\cite{Katsnelson06, Tworzydlo06} This setup was studied experimentally in Refs.\
\onlinecite{Morpurgo06,Miao07,Danneau07, Danneau08} and the ballistic value
$e^2/\pi h$ was indeed observed for large aspect ratios. This geometry of
samples is particularly advantageous for the analysis of evolution from the
ballistic to diffusive transport.

A complete description of the electron transport through a finite system
involves not only the conductance but also higher cumulants of the distribution
of transferred charge. The second moment is related to the current noise in the
system. The intensity of the shot noise is characterized by the Fano factor $F$.
For clean graphene, this quantity was studied in Ref.\ \onlinecite{Tworzydlo06}.
Surprisingly, in a short and wide sample ($W\gg L$) the Fano factor takes
the universal value $F = 1/3$, that coincides with the well-known result for a
diffusive metallic wire. \cite{BeenakkerRMP} This is at odds with usual clean
metallic systems, where the shot noise is absent ($F=0$). The Fano factor
$F=1/3$ in clean graphene is attributed \cite{Tworzydlo06} to the fact that the
current is mediated by evanescent rather than propagating modes. Furthermore,
the whole distribution of transmission eigenvalues for the massless Dirac
equation in a clean sample with $W \gg L$ at the Dirac point agrees with that of
mesoscopic metallic wires in the diffusive regime. \cite{Ludwig07}

The effect of disorder on the shot noise was studied numerically in Refs.\
\onlinecite{San-Jose07, Lewenkopf08}, where the value of the Fano factor $F
\approx 0.3$ was found across the whole crossover form ballistics to diffusion.
The Fano factor close to $1/3$ was also observed at the Dirac point
experimentally. \cite{Danneau07, Danneau08} When the chemical potential was
shifted away from the Dirac point, the Fano factor decreased, then showed an
intermediate shoulder at $F\approx 0.15$, and finally approached zero for
largest gate voltages (carrier concentrations).

While both diffusive and clean limits have been addressed analytically, only
numerical and experimental results for the intermediate regime of ballistic
transport through disordered samples have been available so far.\cite{foot-1D}
The aim of this paper is to fill this gap. We develop the analytic theory of
electron transport in disordered graphene in the ballistic geometry ($L\ll W,l$)
and calculate the full statistics of the charge transfer for both zero (the
Dirac point) and large concentration of carriers. We also discuss the crossover
to diffusive regime and construct the overall ``phase diagram'' of transport
regimes.

The structure of the paper is as follows. We begin in Sec.\ \ref{Sec:Tmatrix}
with the introduction of the model and derivation of a general transfer-matrix
equation. In Sec.\ \ref{Sec:clean} we calculate transport properties of a clean
sample. In Sec.\ \ref{Sec:first-correction} the disorder is included in the
lowest order of the perturbation theory. The resummation of leading
higher-order corrections to the counting statistics is performed within the
renormalization group approach in Sec.\ \ref{Sec:RG}. For the case of random
potential we present an evidence in favor of a universal scaling of the
distribution function of transmission coefficients valid at the Dirac point for
samples of arbitrary size (covering both ballistic and diffusive regimes). We
summarize the results and discuss the perspectives in Sec.\ \ref{Sec:summary}.
Technical details are relegated to Appendices \ref{App:oscillations},
\ref{App:twoloop}, and \ref{App:oneloop}.

\section{Transfer-matrix technique}
\label{Sec:Tmatrix}

We start with introducing our model and the general formalism of transfer matrix
technique. For graphene, this approach was employed in Refs.\
\onlinecite{Katsnelson06, Tworzydlo06, Titov06,
CheianovFalko, Titov07, San-Jose07}.

We will adopt the single-valley model of graphene.
More specifically, we will consider scattering of
electrons only within a single valley and neglect intervalley scattering events.
Indeed, a number of experimental results show that the dominant disorder in graphene
scatters electrons within the same valley.
First, this disorder model is supported by the odd-integer
quantization\cite{Novoselov05, Zhang05, Geim07} of the
Hall conductivity, $\sigma_{xy} = (2n+1) 2e^2/h$,
representing a direct evidence\cite{OurQHE} in favor of smooth disorder
which does not mix the valleys. The analysis of weak
localization also corroborates the dominance of intra-valley
scattering\cite{Savchenko}. Furthermore, the observation of the
linear density dependence\cite{Geim07} of graphene conductivity away from the
Dirac point implies that the relevant disorder is due to charged impurities
and/or ripples.\cite{Nomura07, Ando06, Nomura06, Khveshchenko, OurPRB,
footnote-unitary} Due to the long-range character of these types of disorder,
the intervalley scattering
amplitudes are strongly suppressed and will be neglected in our treatment.
Finally, apparent absence of localization at the Dirac point down to very low
temperatures\cite{Novoselov05, Zhang05,Kim}
 can be explained only by some special symmetry of
disorder. The most realistic candidate model is the long-range randomness
which does not scatter between valleys. \cite{OurPRL}

The single-valley massless Dirac Hamiltonian of
electrons in graphene has the form (see, e.g., Ref.\ \onlinecite{RMP07})
\begin{equation}
 H
  = v_0 \bm{\sigma} \mathbf{p} + V(x,y),
 \qquad
 V(x,y)
  = \sigma_\mu V_\mu(x,y).
 \label{ham}
\end{equation}
Here $\sigma_\mu$ (with $\mu = 0,x,y,z$) are Pauli matrices
acting on the electron pseudospin degree of freedom corresponding to the sublattice
structure of the honeycomb lattice, $\bm{\sigma} \equiv \{\sigma_x, \sigma_y\}$,
and the Fermi velocity is $v_0 \approx 10^8$ cm/s.
The random part $V(x,y)$ is in general a $2 \times 2$
matrix in the sublattice space. Below we set
$\hbar = 1$ and $v_0 = 1$ for convenience.

\begin{figure}
 \centerline{\includegraphics[width=0.95\columnwidth]{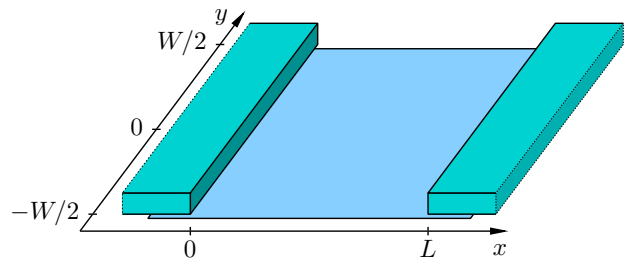}}
 \caption{(Color online) Schematic setup for two-terminal transport
measurements. Graphene sample of dimensions $L \times W$ is placed between two
parallel contacts. We assume $W\gg L$ throughout the paper.}
 \label{Fig:sample}
\end{figure}

We will calculate transport properties of a rectangular graphene sample with
the dimensions $L \times W$. The contacts are attached to the two sides of the
width $W$ separated by the distance $L$. We fix the $x$ axis in the direction of
current, Fig.\ \ref{Fig:sample}, with the contacts placed at $x=0$ and $x=L$.
We assume $W \gg L$, which allows us to neglect the boundary effects related to
the edges of the sample that are parallel to the $x$ axis (at $y=\pm W/2$).

Following Ref.\ \onlinecite{Tworzydlo06}, metallic contacts are modelled as
highly doped graphene regions described by the same Hamiltonian (\ref{ham}). In
other words, we assume that the chemical potential $E_F$ in the contacts is
shifted far from the Dirac point. In particular, $E_F \gg \epsilon$, where
$\epsilon$ is the chemical potential inside the graphene sample counted from the
Dirac point. (All our results are independent of the sign of energy, thus we
assume $\epsilon > 0$ throughout the paper.) A large number of propagating modes
exists in the leads, all belonging to the circular Fermi surface of radius $p_F
= E_F/v_0$. These modes are labelled by the momentum $p_n = 2\pi n/W$ in $y$
direction with $|n| < N = W p_F/2\pi$. Particular boundary conditions at $y =
\pm W/2$ shift the quantized values of $p_n$ by a constant of order $1/W$.
However, this constant has no significance in the limit $W \gg L$ when many
channels participate in electron transport.

Clearly, the transverse momentum $p_n$ is preserved in the clean system. We will
use the  mixed momentum-coordinate representation, with the wave function
$\Psi_n(x)$ bearing a vector index $n$ in the space of transverse momenta
supplemented by a 2-spinor structure in pseudospin (sublattice) space. The
eigenstates of the clean Hamiltonian $H_0=v_0\bm{\sigma} \mathbf{p}$ have the
direction of pseudospin  parallel to the electron momentum. It is convenient to
perform the unitary rotation\cite{Titov07} in the pseudospin space $\psi =
\mathcal{L} \Psi$ with $\mathcal{L} = (\sigma_x + \sigma_z)/\sqrt{2}$ which
transforms $\sigma_x$ to the diagonal form: $\mathcal{L} \sigma_x
\mathcal{L}^\dagger = \sigma_z$. Hence the two components of the rotated spinor
correspond to right- and left-propagating waves, \cite{footnote-EFinLeads} $\psi
= \{ \psi^R, \psi^L \}$. In terms of the new function $\psi_n(x)$, the
Schr\"odinger equation $H \Psi = \epsilon \Psi$ acquires the form
\cite{San-Jose07, Titov07}
\begin{equation}
 \frac{\partial \psi_n}{\partial x}
  = (\sigma_x p_n +i \sigma_z \epsilon) \psi_n
    -i\sigma_z \sum_m U_{nm}(x) \psi_m.
 \label{psi_evolution}
\end{equation}
The matrix $U_{nm}(x)$ represents the operator $\mathcal{L} V(x,y)
\mathcal{L}^\dagger$ in the mixed momentum-coordinate representation
\begin{equation}
 U_{nm}(x)
  = \int \frac{dy}{W}\, e^{-i(p_n - p_m) y}
    \mathcal{L} V(x,y) \mathcal{L}^\dagger.
\end{equation}

A standard description of electron propagation involves the scattering matrix
$\mathcal{S}$. This is a unitary matrix relating the amplitudes of incident and
outgoing waves
\begin{equation}
 \begin{pmatrix}
  \psi^L(0) \\
  \psi^R(L)
 \end{pmatrix}
  = \mathcal{S} \begin{pmatrix}
      \psi^R(0) \\
      \psi^L(L)
    \end{pmatrix},
 \qquad
 \mathcal{S}
  = \begin{pmatrix}
     r & t' \\
     t & r'
    \end{pmatrix}.
\end{equation}
The elements $t$, $t'$ and $r$, $r'$ are matrices in channel space formed by
transmission and reflection amplitudes, respectively. The unitarity condition
$\mathcal{S}^\dagger \mathcal{S} = \mathbb{1}$ ensures conservation of particle
number.

A closely related formulation is based on the transfer matrix $\mathcal{T}$
which expresses the waves at the point $x=L$ through those at $x=0$:
\begin{equation}
 \begin{pmatrix}
  \psi^R(L) \\
  \psi^L(L)
 \end{pmatrix}
  = \mathcal{T} \begin{pmatrix}
      \psi^R(0) \\
      \psi^L(0)
    \end{pmatrix},
 \quad
 \mathcal{T}
  = \begin{pmatrix}
     {t^\dagger}^{-1} & r' {t'}^{-1} \\
     -{t'}^{-1} r & {t'}^{-1}
    \end{pmatrix}.
 \label{Tdefinition}
\end{equation}
This description is convenient due to the simple multiplicativity property:
$\mathcal{T}(x_3, x_2) \mathcal{T}(x_2, x_1) = \mathcal{T}(x_3, x_1)$.
The current conservation is provided by the identity $\mathcal{T}^\dagger
\sigma_z \mathcal{T} = \sigma_z$.

By definition, the transfer matrix $\mathcal{T}(x_2, x_1)$ yields a solution
to the Schr\"odinger equation (\ref{psi_evolution}) in the form $\psi(x_2) =
\mathcal{T}(x_2, x_1) \psi(x_1)$. Transfer matrix itself, as a function of
its first argument, obeys the same Schr\"odinger equation with the initial
condition $\mathcal{T}(x,x) = \mathbb{1}$. In a clean sample the solution
depends only on the difference $x_2 - x_1$ and is diagonal in channel space:
\begin{equation}
 \mathcal{T}^{(0)}_{nm}(x_2,x_1)
  = \delta_{nm}
    \exp\bigl[ (\sigma_x p_n + i \sigma_z \epsilon) (x_2-x_1) \bigr].
 \label{T0}
\end{equation}

In order to include disorder as a perturbation, it is convenient to cast the
Schr\"odinger equation (\ref{psi_evolution}) into an integral form. In terms of
transfer matrix the integral equation reads
\begin{multline}
 \mathcal{T}(x_2, x_1)
  = \mathcal{T}^{(0)}(x_2, x_1) \\
    -i \int_{x_1}^{x_2} dx\,
      \mathcal{T}^{(0)}(x_2, x) \sigma_z U(x) \mathcal{T}(x, x_1).
 \label{evol}
\end{multline}

The transport statistics of the sample is expressed in terms of transmission
eigenvalues $T_n$ --- the eigenvalues of the matrix $t^\dagger t$. One can
extract these transmission eigenvalues from the upper left element of the
transfer matrix (\ref{Tdefinition}). The first two moments of the transferred
charge distribution determine the conductance (by Landauer formula) and the Fano
factor \cite{BeenakkerRMP}
\begin{equation}
 G
  = \frac{4e^2}{h} \mathop{\mathrm{Tr}}(t^\dagger t),
 \qquad
 F
  = 1 - \frac{\mathop{\mathrm{Tr}}(t^\dagger t)^2}{\mathop{\mathrm{Tr}}(t^\dagger t)}.
 \label{GFn}
\end{equation}
The factor $4$ in the expression for the conductance accounts for the spin and
valley degeneracy.

\section{Clean graphene}
\label{Sec:clean}

We will first analyze transport properties of a clean graphene strip. In the
``short and wide'' geometry ($W\gg L$) we are considering, the total number of
channels 
participating in charge transfer is large. This allows us to replace
summation over channels by integration. From now on, we will identify
channels by the dimensionless momentum $p=p_n L$ in $y$ direction and integrate
over this 
momentum according to
\begin{equation}
 \sum_n
  \mapsto \frac{W}{L} \int \frac{dp}{2\pi}.
 \label{measure}
\end{equation}

The transfer matrix $\mathcal{T}^{(0)}$, and hence its upper-left block
${t^\dagger}^{-1}$, are diagonal in channels. Using the explicit form of the
clean graphene transfer matrix, Eq.\ (\ref{T0}), one calculates the transmission
eigenvalues \cite{Titov07}
\begin{equation}
 T_p
  = (t^\dagger t)_{pp}
  = \left[
      1 + \frac{p^2 \sinh^2 \sqrt{p^2 - (\epsilon L)^2}}{p^2 - (\epsilon L)^2}
    \right]^{-1}.
 \label{Tclean}
\end{equation}

For the conductance and Fano factor we obtain from Eq.\ (\ref{GFn})
\begin{equation}
 G
  = \frac{2e^2 W}{\pi h L} \int dp\; T_p,
 \qquad
 F
  = 1 - \frac{\int dp\; T_p^2}{\int dp\; T_p}.
 \label{GF}
\end{equation}
The result of numerical integration of Eq.\ (\ref{GF}) is shown in Fig.\
\ref{Fig:GF}.
A detailed
analytical analysis of the two limiting cases of small and large energies is
presented below.

\begin{figure}
 \centerline{\includegraphics[width=0.95\columnwidth]{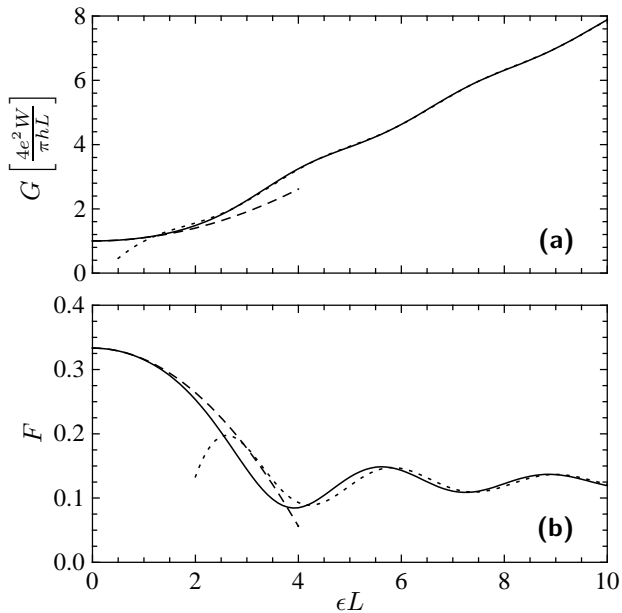}}
 \caption{Energy dependence of the (a) conductance and the (b) Fano factor of
the clean sample with $W \gg L$. Solid lines show numerical results. Low energy
asymptotics Eq.\ (\protect\ref{GFsmallE}) is plotted by dashed lines while
dotted lines correspond to high energy limit Eqs.\ (\protect\ref{Gosc}) and
(\protect\ref{Fosc}). Asymptotical curves provide a very good approximation to
the exact result in the whole range of energies.}
 \label{Fig:GF}
\end{figure}

\subsection{Transmission distribution and counting statistics}
\label{Sec:trans-distr-count-stat}

It is convenient to introduce the distribution function $P(T)$ of transmission
eigenvalues (\ref{Tclean}). This distribution function provides a measure in the
space of channels which is, by definition, equivalent to the integration measure
(\ref{measure}),
\begin{equation}
 P(T)\, dT
  = \frac{W\, d|p|}{\pi L}.
 \label{PTdef}
\end{equation}
According to (\ref{Tclean}), there is one-to-one correspondence between the
transmission eigenvalue $0\le T \le 1$ and the absolute value $|p|$ of the
momentum; an extra factor of $2$ in the right-hand side of
Eq.\ (\ref{PTdef}) accounts for the double degeneracy between channels with
momenta $p$ and $-p$. 

Very generally, the distribution function $P(T)$ determines the full statistics
of the charge transfer. Specifically, one defines the counting statistics
$\kappa(\chi)=\sum_N e^{i\chi N} \mathcal{P}(N)$, where $\mathcal{P}(N)$ is the
probability that $N$ particles are transferred within a measurement time
interval $t_m$. Then $\ln\kappa(\chi)$ is the generating function for cumulants,
\begin{equation}
 \label{cumulants-generating-function}
\ln\kappa(\chi) = \sum_k \frac{(i\chi)^k}{k!} \langle\langle
N^k\rangle\rangle.
\end{equation}
It can be related to $P(T)$ in the following way \cite{LeeLevitovYakovets} (we
assume zero temperature and retain the factor of 4 taking into account the spin
and valley degeneracy in graphene):
\begin{equation}
 \label{counting-statistics-transmissions}
\ln\kappa(\chi) = \frac{4e^2}{h}V t_m \int dT P(T) \ln
(1-T+e^{i\chi}T),
\end{equation}
where $V$ is the applied voltage. In particular, the first two of the cumulants
$\langle\langle N^k\rangle\rangle$ determine the conductance $G$ and the shot
noise power $S$ via $\langle\langle N \rangle\rangle \equiv \langle N \rangle =
V t_m G$ and $\langle\langle N^2 \rangle\rangle = V t_m S$. According to Eqs.\
(\ref{cumulants-generating-function}),
(\ref{counting-statistics-transmissions}), one has
\begin{equation}
\label{GS}
 G
  = \frac{4e^2}{h} \int dT\, T P(T),
 \qquad
 S
  = \frac{4e^2}{h} \int dT\, T(1-T) P(T),
\end{equation}
and the Fano factor
\begin{equation}
 F
  = \frac{S}{G}
  = 1 - \frac{\int dT\, T^2 P(T)}{\int dT\, T P(T)}.
 \label{GFPT}
\end{equation}
The relations (\ref{counting-statistics-transmissions}), (\ref{GS}),
(\ref{GFPT}) are of general validity and equally applicable to the clean and
disordered system. All the information about scattering, both at the interface
with leads and in the bulk of the system, is encoded in the transmission
distribution $P(T)$.  Clearly, Eqs.\ (\ref{GS}), (\ref{GFPT}) for the
conductance and the shot noise are equivalent to Eq.\ (\ref{GFn}).

\subsection{Low energies: $\epsilon L \ll 1$}
\label{Sec:CleanLowEnergies}

In the low energy limit, we calculate the distribution function $P(T)$ in the
form of a power series in the small parameter $\epsilon L$. In order to perform
this calculation, we first invert the function $T_p$ given by Eq.\
(\ref{Tclean}) keeping  terms of the second order in $\epsilon L$:
\begin{gather}
 p(T)
  = p_0(T) + \frac{(\epsilon L)^2}{2} \left[
      \frac{1}{p_0(T)} - \frac{\sqrt{1-T}}{p_0^2(T)}
    \right], \\
 p_0(T)
  = \mathop{\mathrm{arccosh}}\frac{1}{\sqrt{T}}.
\end{gather}
Now we substitute this expression into Eq.\ (\ref{PTdef}) and obtain the
distribution
\begin{multline}
 P(T)
  = \frac{W}{\pi L}\, \frac{dp(T)}{dT} \\
  = \frac{W}{2\pi L} \frac{1}{T \sqrt{1-T}} \left[
      1 + (\epsilon L)^2 \left(
        \frac{\sqrt{1 - T}}{p_0^3(T)}
        - \frac{1 + T}{2 p_0^2(T)}
      \right)
    \right].
 \label{PTe}
\end{multline}
It is worth noticing that by definition $\int dT P(T)$ should give the total
number of open channels $Wp_F/\pi$ in the leads. In fact, the logarithmic
divergence at $T\to 0$ of the normalization of Eq.\ (\ref{PTe}) is cut off at
the lowest transmission eigenvalue $T_{\rm min}\sim \exp(-2p_FL)$. This
small-$T$ cutoff is, however, immaterial for the calculation of the moments
(conductance, noise, etc).

At zero energy, the function $P(T)$ reproduces the well-known Dorokhov result
\cite{Dorokhov83} for a diffusive wire. This is, in particular, the reason for
the $1/3$ Fano factor in graphene. \cite{Tworzydlo06} The fact that the clean
graphene sample is characterized by exactly the same form of the transmission
distribution as a generic diffusive wire is highly nontrivial. We will show
below (Secs.\ \ref{Sec:first-correction} and \ref{Sec:RG}) that this remarkable
correspondence remains valid in the ballistic regime when leading disorder
effects are incorporated.

Using the distribution (\ref{PTe}), we obtain the following results for the
conductance and the Fano factor of clean graphene at low energies, $\epsilon L
\ll 1$,
\begin{align}
 G
  &=\frac{4 e^2}{\pi h}\; \frac{W}{L} \left[
     1 + c_1 (\epsilon L)^2
   \right],
 \quad
 F
  =\frac{1}{3} \left[
     1 + c_2 (\epsilon L)^2
   \right],
\label{GFsmallE}\\
 c_1
  &=\frac{35 \zeta(3)}{3 \pi^2} - \frac{124 \zeta(5)}{\pi^4}
  \approx 0.101,
 \label{c1}\\
 c_2
  &=-\frac{28 \zeta(3)}{15 \pi^2} - \frac{434 \zeta(5)}{\pi^4}
    +\frac{4572 \zeta(7)}{\pi^6}
  \approx -0.052.
 \label{c2}
\end{align}
At the Dirac point ($\epsilon=0$), Eq.\ (\ref{GFsmallE}) reproduces the earlier
analytical results of Refs.\ \onlinecite{Katsnelson06, Tworzydlo06}. Low energy
asymptotics is shown with dashed lines in Fig.\ \ref{Fig:GF}.

\subsection{High energies: $\epsilon L \gg 1$}
\label{Sec:clean-high-energy}

When the Fermi energy $\epsilon$ in the sample is far from the Dirac point,
many conducting ($T\sim 1$) channels are opened. In this regime, the
conductivity and higher moments of the transmission distribution are essentially
linear in $\epsilon$ with small oscillating corrections (see Fig.\
\ref{Fig:GF}). These oscillations are due to interference effects: conductance
is relatively enhanced and the noise is suppressed when a channel exhibits
resonant transmission with $T$ close to $1$. This phenomenon is similar to the
Fabry-Perot resonances.

We begin with the calculation of the main (proportional to $\epsilon$) part of
the transmission distribution function and will return to the oscillatory
correction later in this section. It is convenient to find first the generating
function of transmission moments, defined as
\begin{equation}
 \mathcal{F}(z)
  = \sum_{n=1}^\infty z^{n-1} \mathop{\mathrm{Tr}} (t^\dagger t)^n
  = \mathop{\mathrm{Tr}} \left[
      t^{-1} {t^\dagger}^{-1} - z
    \right]^{-1}.
 \label{Fdef}
\end{equation}
This function appears to be very useful for the forthcoming calculation of the
transport properties of a disordered sample. In this section, we apply it to the
clean system. According to Eqs.\ (\ref{Fdef}) and (\ref{Tclean}), we have
\begin{equation}
 \mathcal{F}(z)
  = \frac{W}{L} \int \frac{dp}{2\pi} \left[
      1 - z + \frac{p^2 \sin^2 \sqrt{(\epsilon L)^2 - p^2}}
                   {(\epsilon L)^2 - p^2}
    \right]^{-1}.
 \label{F0int}
\end{equation}
The integrand oscillates rapidly in the interval $-\epsilon L < p < \epsilon L$.
This interval of momenta contains all open (well-conducting) channels and thus
provides the main contribution to the generating function. At high energies, it
is convenient to introduce a new variable $u$, such that
\begin{equation}
 p
  = p_n L =\epsilon L\sqrt{1 - u^2}, \label{u-index}
\end{equation}
\begin{eqnarray}
\int dp f(p)
  &\mapsto& \epsilon L \int_0^1 \frac{u\, du}{\sqrt{1-u^2}} \nonumber \\
&\times& \bigl[
      f(\epsilon L\sqrt{1 - u^2}) + f(-\epsilon L\sqrt{1 - u^2})
    \bigr].\nonumber\\
\label{u-mapto}
\end{eqnarray}

Transforming Eq.\ (\ref{F0int}) to the new variable (\ref{u-index}) and
averaging over oscillations (see Appendix \ref{App:oscillations} for
details), we obtain
\begin{multline}
 \mathcal{F}(z)
  = \frac{W \epsilon}{\pi \sqrt{1 - z}}
    \int_0^1 \frac{u^2\, du}{\sqrt{1 - u^2} \sqrt{1 - z u^2}} \\
  = \frac{W \epsilon}{\pi}\, \frac{K(z) - E(z)}{z \sqrt{1-z}}.
 \label{Fz}
\end{multline}
Here $K(m)$ and $E(m)$ are complete elliptic integrals of the first and second
kind \cite{footnote-elliptic} with the parameter $m$.

\begin{figure}
 \centerline{\includegraphics[width=0.6\columnwidth]{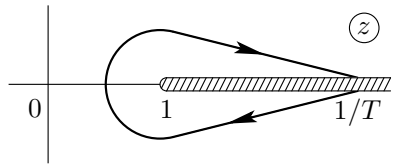}}
 \caption{Integration contour used in Eq.\ (\protect\ref{F-P-integral2}).}
 \label{Fig:RH}
\end{figure}

The function $\mathcal{F}(z)$ is regular at the point $z=0$. The coefficients
of the
series expansion near this point provide the moments of transferred charge
distribution, see Eq.\ (\ref{Fdef}).
The transmission distribution function $P(T)$ is related to $\mathcal{F}(z)$
by the linear integral equation
\begin{equation}
 \int_0^1 \frac{P(T)}{T^{-1} - z}\, dT
  = \mathcal{F}(z),
 \label{F-P-integral}
\end{equation}
which follows from Eq.\ (\ref{Fdef}). In order to solve this equation for
$P(T)$, we note that the function $\mathcal{F}(z)$ has a branch cut along real
axis running from $1$ to $\infty$. We integrate Eq.\ (\ref{F-P-integral}) along
a contour going from $z = 1/T - i0$ to $z = 1/T + i0$ encircling the point $z =
1$, see Fig.\ \ref{Fig:RH}. This integration yields
\begin{multline}
 \int\limits_{1/T - i0}^{1/T + i0} \frac{dz}{2\pi i}\, \mathcal{F}(z)
  = \int_0^1 P(\tilde{T}) d\tilde{T}
    \int\limits_{1/T - i0}^{1/T + i0} \frac{dz}{2\pi i ({\tilde{T}}^{-1} - z)}
\\
  = \int_T^1 P(\tilde{T}) d\tilde{T}.
 \label{F-P-integral2}
\end{multline}
To find the distribution function, we calculate the derivative of the above
equation with respect to $T$ and obtain
\begin{equation}
 P(T)
  = \frac{\mathcal{F}(1/T + i0) - \mathcal{F}(1/T - i0)}{2\pi i T^2}.
 \label{P-F}
\end{equation}
This identity establishes a relation between the distribution function $P(T)$
and the jump of $\mathcal{F}(z)$ across the branch cut at the point $z = 1/T$.
In other words, equation (\ref{F-P-integral}) solves the corresponding
Riemann-Hilbert problem.

To find the explicit formula for the distribution $P(T)$, we perform an analytic
continuation of the expression (\ref{Fz}) from the vicinity of the point $z = 0$
to $z = 1/T \pm i0$ and substitute the result into Eq.\ (\ref{P-F}). This yields
\begin{equation}
 P(T)
  = \frac{W \epsilon}{\pi^2}\;
    \frac{K(T) - E(T)}{T \sqrt{1 - T}}.
 \label{PTelarge}
\end{equation}
This distribution function provides a full transport description at high
energies (up to small oscillatory corrections discussed below).
We note that Eq.\ (\ref{PTelarge}) does not take into account almost closed
(evanescent) channels with $p_n > \epsilon$ (and thus $T \ll 1$). Estimating
their contribution, we find that it is suppressed by a factor 
$(\epsilon L)^{-4}$ compared to the main term, thus yielding a
negligible contribution to the charge transfer.

Using the above distribution (or, equivalently, the generating function),
we calculate the asymptotics of the conductance and Fano factor in the high-energy
limit $\epsilon L \gg 1$,
\begin{equation}
 G
  = \frac{e^2}{h}\; W \epsilon,
 \qquad\qquad
 F
  = \frac{1}{8}.
 \label{GFelarge}
\end{equation}

We recall that when passing from Eq.\ (\ref{F0int}) to Eq.\ (\ref{Fz}), we
neglected the oscillatory contributions to the generating function. A more
accurate calculation accounting for these oscillations is presented in Appendix
\ref{App:oscillations} [see Eq.\ (\ref{Fzosc})]. The results for the
conductance and the Fano factor read
\begin{align}
 G
  &= \frac{e^2}{h}\; W \epsilon \left[
      1 + \frac{\sin (2\epsilon L - \pi/4)}{2\sqrt{\pi} (\epsilon L)^{3/2}}
    \right], \label{Gosc} \\
 F
  &= \frac{1}{8} \left[
      1 - \frac{9 \sin (2\epsilon L - \pi/4)}{2\sqrt{\pi} (\epsilon L)^{3/2}}
    \right]. \label{Fosc}
\end{align}
These results are in a good agreement with the high-energy
behavior of $G$ and $F$ calculated numerically, see Fig.\ \ref{Fig:GF}.

Let us emphasize that transport properties of the system at high energies
depend on the particular model of the contacts. \cite{Schomerus07, Blanter07}
In our calculation we assume that the boundaries between graphene and the leads
are sharp. This model is well justified if the actual extension $d$ of the
transitional region at the interface is small compared to the electron
wavelength in graphene. This wavelength is energy dependent: it tends to
infinity at $\epsilon=0$ and decreases with increasing $\epsilon$. Thus the
small-energy results, Sec.\ \ref{Sec:CleanLowEnergies}, are universal and not
influenced by the microscopic details of the interface provided the size of the
boundary transitional region is much smaller than the length of the sample $L$.
On the other hand, the results of the current section are applicable for not too
high energies, $\epsilon\ll 1/d$. For higher energy, the electron wavelength
becomes comparable to $d$ and the transmission properties of the sample become
nonuniversal. In the extreme high-energy limit, $\epsilon d \gg 1$ the boundary
becomes adiabatically smooth. This, in particular, leads to the vanishing Fano
factor because the semiclassically propagating electrons are either transmitted
or reflected without any uncertainty.

Our results for the energy dependence of the conductance and the Fano factor in
clean graphene are in agreement with the findings of Refs.\
\onlinecite{Tworzydlo06, Titov07}, where the sum over transmission channels was
evaluated numerically for a finite (but sufficiently large) ratio $W/L$.
Experimentally, such a ballistic setup was studied in Refs.\
\onlinecite{Danneau07, Danneau08}. Most of the experimental observations
reasonably agree with our results. The ``conductivity'' $GL/W$ (which is equal
to $4e^2/\pi h$ at the neutrality point, as expected for a ballistic sample)
increases roughly linearly with energy $\epsilon$. The Fano factor has a value
close to $1/3$ at the Dirac point and decreases when one moves away from the
Dirac point, showing a tendency to saturate at $F\approx 0.15$, which is not
far from the value 1/8 we have obtained in the high-energy regime. Measurements
on other samples reveal that very far from the Dirac point the Fano factor
decreases again, reaching a value as low as 0.02. Apparently, the intermediate
plateau corresponds to the high-energy regime $L^{-1}\ll \epsilon \ll d^{-1}$
investigated in our work, while the vanishing of the Fano factor at still
higher electron concentrations corresponds to the ultra-high-energy range,
$\epsilon \gg d^{-1}$. It is not quite clear to us why the oscillatory
structures are not observed in experimental data. A possible explanation
is that the length $L$ of the sample in the experiment varies as a function of
the $y$ coordinate, leading to a suppression of the oscillations.

\section{Including disorder: Perturbative treatment}
\label{Sec:first-correction}

So far, we have considered the transport properties of a clean graphene sample.
In the present section we include disorder on the level of the leading
perturbative correction. As discussed in the beginning of Sec.\
\ref{Sec:Tmatrix}, we neglect the intervalley scattering. Further, we will
assume the Gaussian statistics for disorder components $V^\mu$ that determine
the random part $V(x,y)$ of the Hamiltonian (\ref{ham}) acting within a single
valley,
\begin{gather}
 \langle V^\mu(x, y) \rangle
  = 0, \\
 \begin{gathered}
  \langle V^\mu(x, y) V^\nu(x', y') \rangle
   = 2\pi \delta_{\mu\nu} w_\mu(x - x', y - y').
 \end{gathered}
 \label{Gauss}
\end{gather}
This type of randomness
is realized when the scattering is due to impurities in the substrate separated
by a thick (compared to the lattice constant) clean spacer layer from the
graphene plane. The intervalley matrix elements of the disorder potential are
then exponentially suppressed and can be safely neglected. A more realistic case
of long-range charged impurities with $1/r$ potentials can also be treated
perturbatively within the Gaussian model, but with an energy-dependent
scattering amplitude. \cite{OurPRB} We will briefly discuss modifications of the
results in the case of Coulomb-type impurities in Sec.\
\ref{Sec:additional_comments}.

The correlation function $w_\mu(x,y)$ is even with respect to both arguments and
is peaked at short (compared to the wavelength in the sample) distances, being
hence almost a delta function. At the same time, we will have to keep a small
but non-zero correlation length in order to regularize ultraviolet singularities
arising at the intermediate stage of our calculation. The results of the
calculation will not depend on this correlation length.

In the transfer-matrix approach, it is convenient to convert the correlation
function to momentum representation in $y$ direction. The $x$ dependence of
$w_\mu$ can be safely replaced by the delta function without generating any
singularities. Thus we introduce new dimensionless functions $\alpha_\mu$
according to
\begin{equation}
 w_\mu(x,y)
  = \frac{\delta(x)}{W} \sum_m e^{i q_m y}\; \alpha_\mu(q_m L).
 \label{corr}
\end{equation}

The functions $\alpha_\mu(q)$ vary slowly with $q$. They are almost constant at
low values of momentum and decay at the large scale of inverse correlation
length. We will express the transport characteristics of the system in terms of
four constants
\begin{equation}
 \alpha_\mu
  = \alpha_\mu(0).
\end{equation}
These parameters are nothing but the amplitudes of the effective delta
functions in Eq.\ (\ref{Gauss}),
\begin{equation}
 w_\mu(x,y)
  \approx \alpha_\mu \delta(x) \delta(y),
 \label{alpha-mu}
\end{equation}
and correspond to the intravalley scattering parameters used in
Ref.\ \onlinecite{OurPRL} (there it was assumed that
$\alpha_x=\alpha_y\equiv \alpha_\perp/2$).

In the present section we will calculate the first disorder correction to the
transport properties of a graphene sample. Specifically, we will find a
linear-in-$\alpha_\mu$ contribution to the function $P(T)$.
It is convenient to introduce short-hand notations for inverse transmission
amplitudes and probabilities of the clean sample,
\begin{equation}
 h_n
  = 1/t_n^{(0)},
 \qquad
 H_n
  = |h_n|^2
  = 1/T_n^{(0)}.
 \label{hH}
\end{equation}
Further, it will be useful to label the types of disorder ($\mu=0,x,y,z$) by a
pair of binary indices $\xi,\eta=\pm$, according to
\begin{equation}
\begin{aligned}
 \alpha_0
  &= \alpha_{++}~,
 \qquad
 \alpha_x
  = \alpha_{+-}~, \\
 \alpha_y
  &= \alpha_{--}~,
 \qquad
 \alpha_z
  = \alpha_{-+}~.
\end{aligned}
\label{vocabulary}
\end{equation}

We develop the perturbative expansion by iteratively solving Eq.\ (\ref{evol}).
Then we single out the upper-left block of the matrix $\mathcal{T}(L,0)$ thus
obtaining ${t^\dagger}^{-1}$. Up to the second order in $V$ the result is
\begin{equation}
 \bigl( {t^\dagger}^{-1} \bigr)_{mn}
  = \delta_{mn} h_n^*  + \Delta_{mn} - \delta_{mn} h_n^* \sum_l A_{ml}.
 \label{t_inverse}
\end{equation}
Here $\Delta_{mn}$ is the linear correction to the transfer matrix,
\begin{equation}
 \Delta
  = -i \int_0^L dx \left[
      \mathcal{T}^{(0)}(L,x) \sigma_z U(x) \mathcal{T}^{(0)}(x,0)
    \right]_{1,1},
 \label{Delta}
\end{equation}
where the subscript
$1,1$ refers to the upper-left block in the right/left-mover space,
see Eq.\ (\ref{Tdefinition}).

The last term in Eq.\ (\ref{t_inverse}) represents the contribution of the
second order in disorder amplitudes $U(x)$. Since we are interested in the
correction to transport coefficients of the linear order in $\alpha_\mu$ (and
thus quadratic in $U$), we can perform disorder averaging of this term using
Eqs.\ (\ref{Gauss}), (\ref{corr}). Then the integration over $x$-coordinate in
this term is trivial due to the delta function in the correlator (\ref{corr})
and the multiplicativity property of the transfer matrix. We have also used the
relation $\int_0^\infty dx\, \delta(x) = 1/2$. [This identity holds because the
delta function in Eq.\ (\ref{corr}) is a replacement for some symmetric sharply
peaked function.] As a result, we get
\begin{equation}
 A_{mn}
  = \frac{\pi L}{W} \sum_{\xi,\eta = \pm}
    \xi\, \alpha_{\xi\eta}(q_m L - q_n L).
 \label{A}
\end{equation}
The sum over intermediate states $l$ in the last term of Eq.\ (\ref{t_inverse})
converges due to a non-zero correlation length of disorder, encoded in the
momentum dependence of $\alpha_{\xi\eta}$ in Eq.\ (\ref{A}).

Now we substitute the expression (\ref{t_inverse}) and its Hermitian conjugate
into Eq.\ (\ref{Fdef}) and then expand $\mathcal{F}(z)$ up to the second order
in $\Delta$ and first order in $A$. Performing the disorder averaging of terms
containing $\Delta$, we obtain the following expression for the generating
function $\mathcal{F}(z)$:
\begin{multline}
 \mathcal{F}(z)
  = \sum_n \frac{1}{H_n - z} \\
    + \sum_{mn} \left[
      \frac{2 H_n A_{mn}}{(H_n - z)^2}
       + \frac{2 B_{mn} + (H_n + z) C_{mn}}{(H_n - z)^2 (H_m - z)}
    \right]
 \label{Fdisord}
\end{multline}
with
\begin{equation}
 B_{mn}
  = \mathop{\mathrm{Re}} h_m h_n \langle \Delta_{mn} \Delta_{nm}\rangle,
 \quad
 C_{mn}
  = \langle |\Delta_{mn}|^2 \rangle.
 \label{BC}
\end{equation}
In a general case, the two matrices $B_{mn}$ and $C_{mn}$ are very complicated
functions of $m$ and $n$. We will simplify further analysis by considering two
limiting cases of low and high energy.

\subsection{Low energies: $\epsilon L \ll 1$}

We have already calculated the lowest order correction to the distribution
function $P(T)$ due to small energy [see Eq.\ (\ref{PTe})]. Now we are going to
find the lowest disorder correction at exactly zero energy. To the main order,
these two contributions merely add up.

At zero energy, there are no propagating modes in graphene, all the channels are
evanescent. In this situation, it is convenient to use the transverse momentum
$p=p_n L$ instead of index $n$ to label the channels according to Eq.\
(\ref{measure}). The bare transfer matrix (\ref{T0}) at $\epsilon = 0$
simplifies to
\begin{equation}
 \mathcal{T}^{(0)}_p
  = \begin{pmatrix}
     \cosh p & \sinh p \\
     \sinh p & \cosh p
    \end{pmatrix}.
\end{equation}
The quantities $h_n$ and $H_n$ introduced in Eq.\ (\ref{hH}) take the form
\begin{equation}
 h_p
  = \cosh p,
 \qquad
 H_p
  = \cosh^2 p.
\end{equation}
The matrix $\Delta_{nm}$ defined by Eq.\ (\ref{Delta}) now becomes
\begin{multline}
 \Delta_{pq}
  = \int_0^L dx \Bigl\{
      -i V^0_{pq} (x) \cosh[p - (p + q)x/L]\\
      -i V^x_{pq} (x) \cosh[p - (p - q)x/L]\\
      +  V^y_{pq} (x) \sinh[p - (p - q)x/L]\\
      +i V^z_{pq} (x) \sinh[p - (p + q)x/L]
    \Bigr\}.
\end{multline}

Two types of averages [Eq.\ (\ref{BC})] arise in the calculation of transport
properties. These averages are the result of applying Eq.\ (\ref{Gauss}) to the
product of two $\Delta_{pq}$ matrices and subsequent integration over single
[due to the delta function in Eq.\ (\ref{corr})] position $x$.
This yields
\begin{align}
 B_{pq}
  &= -\frac{\pi L}{W} h_p h_q
     \sum_{\xi,\eta = \pm} \alpha_{\xi\eta}(p-q) \biggl[
      \xi \cosh(p - \eta q) \notag \\
      &\hspace{4cm} + \frac{\sinh(p + \eta q)}{p + \eta q}
    \biggr], \label{B} \\
 C_{pq}
  &= \frac{\pi L}{W} \sum_{\xi,\eta = \pm} \alpha_{\xi\eta}(p-q) \left[
      \xi + \frac{\sinh 2 p + \eta \sinh 2 q}{2(p + \eta q)}
    \right]. \label{C}
\end{align}

Now we substitute Eqs.\ (\ref{A}), (\ref{B}), and (\ref{C}) into Eq.\
(\ref{Fdisord}) and separate the resulting expression into four parts,
\begin{align}
 \mathcal{F}(z)
  &= \mathcal{F}_0(z) + \mathcal{F}_1(z) + \mathcal{F}_2(z) + \mathcal{F}_3(z),
\\
 \mathcal{F}_0(z)
  &= \frac{W}{2\pi L} \int \frac{dp}{H_p - z}, \\
 \mathcal{F}_1(z)
  &= \frac{W}{2\pi L} \sum_{\xi,\eta} \int dp\, dq\;
    \xi \eta\, \alpha_{\xi\eta}(p - q) \notag \\
    &\qquad \times \frac{H_p H_q \tanh p \tanh q}{(H_p - z)^2(H_q - z)},
    \label{F1}\\
 \mathcal{F}_2(z)
  &= \frac{W}{4\pi L} \sum_{\xi,\eta} \xi\, \alpha_{\xi\eta}
    \!\int\! dp\, dq\; \frac{z + (1 - 2z)H_p}{(H_p - z)^2(H_q - z)},
\label{F2} \\
 \mathcal{F}_3(z)
  &= -\frac{W}{8\pi L} \sum_{\xi,\eta} \alpha_{\xi\eta}
    \int \frac{dp\, dq}{p + \eta q} \notag\\
    &\qquad \times \left[
      \frac{\sinh 2 p}{(H_p - z)^2} + \frac{\eta \sinh 2 q}{(H_q - z)^2}
    \right].
  \label{F3}
\end{align}
The first part, $\mathcal{F}_0$, originating from the first term of Eq.\
(\ref{Fdisord}), is the generating function for the clean sample,
\begin{equation}
\mathcal{F}_0(z)
  = \frac{W}{\pi L}\, \frac{\arcsin\sqrt{z}}{\sqrt{z - z^2}}.
\label{F0z}
\end{equation}
It corresponds to the distribution function $P(T)$, Eq.\ (\ref{PTe}), at
$\epsilon = 0$.

The other three terms are disorder-induced corrections. The integral in Eq.\
(\ref{F1}) would not be absolutely convergent if we replace $\alpha_{\xi\eta}$
by constants. For this reason, we have to retain the momentum dependence of
$\alpha_{\xi\eta}$ (originating from finite correlation length of disorder) in
the integrand. Performing first the integration over $p + q$ and then over $p -
q$, we get
\begin{equation}
 \mathcal{F}_1(z)
  = \mathcal{F}_0(z) \sum_{\xi,\eta} \xi\eta\, \alpha_{\xi\eta}.
\label{F1z}
\end{equation}
Note that the value of $\mathcal{F}_1$ does not actually depend on the
precise form of the functions $\alpha_{\xi\eta}(p-q)$, but only on their
values at $p-q=0$. Indeed, the integral over $p + q$ and the subsequent integral
over $p - q$ are convergent even with constant $\alpha_{\xi\eta}$. The
finite disorder correlation length is needed only to ensure the absolute
convergence of the $q$ integral in Eq.\ (\ref{F1}).

The integral in $\mathcal{F}_2$, Eq.\ (\ref{F2}), is absolutely convergent in
both variables. This allows us to neglect the momentum dependence of
$\alpha_{\xi\eta}(p-q)$ and replace it by a constant from the very beginning,
yielding
 \begin{equation}
\mathcal{F}_2(z)
  = \mathcal{F}_0(z)\sum_{\xi,\eta} \xi\, \alpha_{\xi\eta}.
\label{F2z}
\end{equation}
The last term $\mathcal{F}_3$ is also absolutely convergent in both
variables. When writing Eq.\ (\ref{F3}), we have simplified the integrand
by means of symmetrization with respect to $p \leftrightarrow \eta q$.
The integrand in Eq. (\ref{F3}) can be rewritten in the form of a total
derivative,
\begin{equation}
 \left(
   \frac{\partial}{\partial p} - \eta\, \frac{\partial}{\partial q}
 \right) \frac{H_p - H_q}{(p+ \eta q)(H_p - z)(H_q - z)},
\end{equation}
and hence
\begin{equation}
\mathcal{F}_3(z)=0.
\label{F3z}
\end{equation}

Collecting all the terms, we finally obtain the following generating function
\begin{equation}
 \mathcal{F}(z)
  = (1 + 2\alpha_0 - 2\alpha_z)\,
    \frac{W}{\pi L}\, \frac{\arcsin\sqrt{z}}{\sqrt{z - z^2}}.
 \label{Falpha}
\end{equation}
We see that the random vector potential, $\alpha_{x,y}$, does not influence
transport characteristics of the system in the lowest order. In fact, any vector
potential is unable to alter conductance or higher moments of charge
transmission at zero energy. We will give a general proof of this statement in
Sec.\ \ref{Sec:vector-potential} below. Another manifestation of this property
was found in Refs.\ \onlinecite{OurPRB, Ludwig, Tsvelik} where it was shown that
the random vector potential does not change the conductivity of an infinite
graphene sample at the Dirac point.

The distribution of transmission eigenvalues follows from Eq.\ (\ref{Falpha})
with the help of identity (\ref{P-F}). Together with the energy
correction from Eq.\ (\ref{PTe}), $P(T)$ acquires the form
\begin{multline}
 P(T)
  = \frac{W}{2\pi L} \frac{1}{T \sqrt{1-T}} \biggl[
      1 + 2(\alpha_0 - \alpha_z) \\
      +(\epsilon L)^2 \left(
        \frac{\sqrt{1 - T}}{\mathop{\mathrm{arccosh}}^3(1/\sqrt{T})}
        -\frac{1 + T}{2 \mathop{\mathrm{arccosh}}^2(1/\sqrt{T})}
      \right)
    \biggr].
 \label{PTUB}
\end{multline}
Remarkably, the functional dependence $P(T)$ is not changed by disorder at
$\epsilon = 0$. We will discuss the consequences of this fact in Sec.\
\ref{Sec:single-parameter-scaling}.

\subsection{High energies: $\epsilon L \gg 1$}
\label{high-ballistic}

The transport properties of a clean graphene sample at high energies were
considered in Sec.\ \ref{Sec:clean-high-energy}. The main contribution to the
conductance and to higher moments is proportional to $\epsilon L$ and comes from
the band of fully opened channels with $|p_n| < \epsilon$. In the present
section we will calculate the disorder-induced correction coming from the same
channels. As we will show below, the relative correction is of the order of
$\alpha_\mu \epsilon L$. Since all the momentum integrals will be restricted to
$|p_n| < \epsilon$, we do not need the ultraviolet regularization and can
neglect the momentum dispersion of $\alpha_\mu$ from the very beginning.

As appropriate for high energies (see Sec.\ \ref{Sec:clean-high-energy}),
we will label the channels
by variables $u$ and $v$ related to $p_n$ and $p_m$
according to Eq.\ (\ref{u-index}). In this representation the quantities $h_n$
and $H_n$ introduced in Eq.\ (\ref{hH}) take the form
\begin{align}
 h_u
  &= \cos(u \epsilon L) -i\, \frac{\sin(u \epsilon L)}{u}, \\
 H_u
  &= \cos^2(u \epsilon L) + \frac{\sin^2(u \epsilon L)}{u^2}.
\end{align}
The matrix $\Delta_{uv}$ [Eq.\ (\ref{Delta})] and the averages $B_{uv}$ and
$C_{uv}$ [Eqs.\ (\ref{B}) and (\ref{C})] contain rapidly oscillating terms. The
integration over $u$ and $v$ will average out these oscillations. For this
reason, we can drop all the terms in $B_{uv}$ and $C_{uv}$ that are proportional
to odd powers of $\sin(u \epsilon L)$ or $\sin(v \epsilon L)$, already before
calculating the integrals in Eq.\ (\ref{Fdisord}). Furthermore, we discard the
contributions that are odd functions of $p_n$ and/or $p_m$, which corresponds to
dropping odd powers of $\sqrt{1-u^2}$ and $\sqrt{1-v^2}$. The matrices $B_{uv}$
and $C_{uv}$ simplify to
\begin{align}
 B_{uv}
  &= -\frac{\pi L}{W} \sum_{\xi,\eta} \alpha_{\xi\eta} \left[
      \xi H_u H_v + \frac{1}{u^2 v^2}
    \right], \\
 C_{uv}
  &= \frac{\pi L}{W} \sum_{\xi,\eta} \alpha_{\xi\eta} \left[
      \xi + \frac{1}{u^2 v^2}
    \right].
\end{align}

Substituting these expressions together with Eq.\ (\ref{A}) into Eq.\
(\ref{Fdisord}) and averaging over oscillations, we find
\begin{multline}
 \mathcal{F}(z)
  = \frac{W \epsilon}{\pi}
    \int_0^1 \frac{du}{\sqrt{(1 - u^2)(1 - z u^2)}} \Biggl[
      \frac{u^2}{\sqrt{1-z}}\\
      +\epsilon L \sum_{\xi,\eta} \alpha_{\xi\eta}
        \int_0^1 \frac{v^2\; dv\, (\xi u^2 - 1)}{\sqrt{(1-v^2)(1-v^2 z)^3}}
    \Biggr].
\label{F0Fdis}
\end{multline}
The first term in the square brackets gives the generating function of the clean
sample [Eq.\ (\ref{Fz})], while the second term represents the leading
disorder-induced correction, $\mathcal{F}_{\text{dis}}(z)$.
Evaluating integrals in Eq.\ (\ref{F0Fdis}), we express this correction in terms
of elliptic integrals:
\begin{multline}
 \mathcal{F}_{\text{dis}}(z)
  = -\frac{W L \epsilon^2}{\pi z^2 (1 - z)}
    \sum_{\xi,\eta} \xi\, \alpha_{\xi\eta} [(1 - z)K(z) - E(z)]\\
  \times[(1 - \xi z)K(z) - E(z)].
 \label{Fcorr}
\end{multline}
Expanding this generating function at $z = 0$, we readily calculate disorder
corrections to the conductance and Fano factor. Combining these corrections with
the results for clean sample, Eqs.\ (\ref{Gosc}) and (\ref{Fosc}), we obtain
\begin{multline}
 G
  = \frac{e^2}{h}\; W \epsilon \biggl[
      1 + \frac{\sin (2\epsilon L - \pi/4)}{2\sqrt{\pi} (\epsilon L)^{3/2}} \\
      -\frac{\pi}{4}\, \epsilon L (\alpha_0 + \alpha_x + 3\alpha_y + 3\alpha_z)
    \biggr],
 \label{Gelarge}
\end{multline}
\begin{multline}
 F
  = \frac{1}{8} \biggl[
      1 - \frac{9\sin (2\epsilon L - \pi/4)}{2\sqrt{\pi} (\epsilon L)^{3/2}} \\
      +\frac{\pi}{4}\, \epsilon L (
        3\alpha_0 + 3\alpha_x + 13\alpha_y + 13\alpha_z
      )
    \biggr].
 \label{Felarge}
\end{multline}
We see that at high energies any disorder suppresses conductance and enhances
noise at the level of the lowest perturbative correction.

To find the disorder correction to transmission distribution function
(\ref{PTosc}) we perform the analytic continuation of
$\mathcal{F}_{\text{dis}}(z)$ from the vicinity of the point $z = 0$ to $z = 1/T
\pm i0$ and apply Eq.\ (\ref{P-F}). The result is
\begin{multline}
 P_{\text{dis}}(T)
  = \frac{W L \epsilon^2}{2 \pi^2} \sum_{\xi,\eta} \alpha_{\xi\eta} \\
    \times \frac{4 E(T)[K(1-T)-\xi E(1-T)]+\pi(\xi-1)}{1-T}.
 \label{65}
\end{multline}
There is, however, a subtlety in determination of $P_{\text{dis}}(T)$,
which is related to the singularity of $\mathcal{F}_{\text{dis}}(z)$ at $z=1$,
see discussion of a similar problem in the clean case (Appendix
\ref{App:oscillations}). As a result, the distribution function (\ref{65})
cannot be applied in the vicinity of $T = 1$. Specifically, we have to impose
the bound
\begin{equation}
 1 - T
  \gg (\alpha_\mu \epsilon L)^2.
\end{equation}
At $1 - T \sim (\alpha_\mu \epsilon L)^2$, disorder-induced correction
(\ref{65}) becomes comparable to the main (clean) term [Eq.\ (\ref{PTelarge})]
and our perturbative expansion breaks down. It should be stressed that this
peculiarity in the behavior of $P(T)$ near $T = 1$ does not affect the
evaluation of the moments using the generating function $\mathcal{F}(z)$, which
is based on the behavior of the latter in the vicinity of $z = 0$. Indeed, the
disorder-induced correction Eq.\ (\ref{Fcorr}) to $\mathcal{F}(z)$ [and thus to
the moments, Eqs.\ (\ref{Gelarge}) and (\ref{Felarge})] is controlled by the
small parameter $\alpha_\mu \epsilon L \ll 1$. Note that at sufficiently high
energies $\epsilon L \gtrsim 1/\alpha_\mu$, disorder correction to $P(T)$
becomes comparable to the clean result in the whole range of $T$. This implies a
crossover to the diffusive regime, where the perturbative approach developed in
the present section fails, see Sec.\ \ref{Sec:RG}.

\section{Renormalization group and overall phase diagram}
\label{Sec:RG}

In the previous section we have calculated the lowest disorder correction to
transport properties of a ballistic graphene sample. In the present section we
will discuss the resummation of higher-order contributions.

The second-order and all higher terms contain logarithmic divergences and thus
become important when system is still in the ballistic regime, $L\ll l$.
These logarithms are intrinsic for two-dimensional Dirac fermions subjected to
disorder and were extensively studied in various contexts using renormalization
group technique. \cite{Dotsenko, Ludwig, NersesyanTsvelik, Bocquet,
AltlandSimonsZirnbauer, Guruswamy} Application of such a renormalization group
(RG) to disordered graphene was developed in Refs.\ \onlinecite{Aleiner06,
OurPRB}.

The RG deals with the two-dimensional action describing
disordered Dirac fermions,
\begin{equation}
 S[\psi]
  = \int d^{2}x \bigg[
      \bar\psi \bm{\sigma}\nabla \psi
      -i \epsilon \bar\psi \psi
      +\sum_{\mu} \pi\alpha_\mu (\bar\psi \sigma_\mu \psi)^2
    \bigg],
\end{equation}
where $\psi$ and $\bar\psi$ are two-component fermionic (anti-commuting)
fields. The field-theoretical
description of a disordered system involves also some tool to get rid of
diagrams with closed fermionic loops. This is usually either supersymmetry or
replica trick. In both cases, the fields acquire additional structure in
supersymmetric or replica space. Equivalently, on can derive the RG equations by simply
discarding all diagrams that contain fermionic loops, without extending the
fields.

\begin{figure}
 \centerline{\includegraphics[width=0.8\columnwidth]{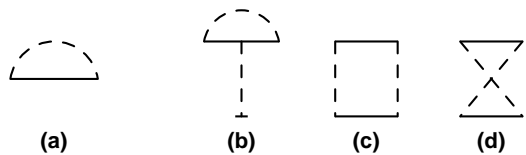}}
 \caption{One-loop diagrams for (a) electron self energy and (b-d) scattering
amplitude. These diagrams yield RG equations (\protect\ref{RGa0}) --
(\protect\ref{RGe}).}
 \label{Fig:oneloop}
\end{figure}

The one-loop diagrams contributing to renormalization of energy and disorder
couplings are shown in Fig.\ \ref{Fig:oneloop}. The solid lines are propagators
of free electrons,
\begin{equation}
 G^{(0)}(\mathbf{p})
  = \frac{\epsilon + \bm{\sigma}\mathbf{p}}{\epsilon^2 - p^2}.
 \label{action}
\end{equation}
Dashed lines stand for disorder correlators $2\pi\sum_\mu \alpha_\mu \sigma_\mu
\otimes \sigma_\mu$. We cut the logarithmic divergence in the one-loop diagrams
by the running scale parameter $\Lambda$, which has the dimension of length, and
obtain the beta functions\ \cite{OurPRB, footnote-normal-dimension}
\begin{align}
 \frac{\partial\alpha_0}{\partial \ln \Lambda}
  &= 2 (\alpha_0 +\alpha_z) (\alpha_0 + \alpha_x + \alpha_y), \label{RGa0}\\
 \frac{\partial\alpha_x}{\partial \ln \Lambda}
  &= \frac{\partial\alpha_y}{\partial \ln \Lambda}
  = 2 \alpha_0 \alpha_z, \label{RGaperp}\\
 \frac{\partial\alpha_z}{\partial \ln \Lambda}
  &= 2 (\alpha_0 + \alpha_z) (-\alpha_z + \alpha_x + \alpha_y),  \label{RGaz}\\
 \frac{\partial\epsilon}{\partial \ln \Lambda}
  &= \epsilon (\alpha_0 + \alpha_x + \alpha_y + \alpha_z).
 \label{RGe}
\end{align}

Bare values of energy and disorder couplings, which are the initial conditions
for RG equations, correspond to the scale of the order of lattice spacing or
disorder correlation length. This scale plays the role of ultraviolet cutoff in
our theory. We will denote it $a$. After renormalization procedure we obtain
renormalized values of the parameters at the scale $\Lambda$ and also a new
effective bandwidth $1/\Lambda$.

The renormalization proceeds until one of the following events happens:
(i) the running scale $\Lambda$ reaches the
system size $L$, (ii) one of the disorder couplings becomes of the order unity,
or (iii) the renormalized energy reaches the bandwidth. We will
discuss these three possibilities for particular disorder types below.
Once the renormalization has been performed, we can calculate observables
by simply applying the perturbation theory.
The results of previous section for transport characteristics
thus remain applicable with bare parameters replaced by their renormalized
values.

\subsection{Random scalar potential}
\label{RSP}

We start the discussion of various disorder types with the case of random scalar
potential.
Let us first consider the zero energy limit when the only parameter of the model is
the disorder coupling $\alpha_0$. In the single-parameter case, the RG beta
function is universal, i.e. does not depend on the regularization scheme, within
the two-loop accuracy. A discussion of the universality and the derivation of the
second-loop contribution is presented in Appendix \ref{App:twoloop}.
The two-loop RG equation reads
\begin{equation}
 \frac{\partial\alpha_0}{\partial \ln \Lambda}
  = 2 \alpha_0^2 + 2 \alpha_0^3.
 \label{RGalpha0}
\end{equation}
The disorder strength, quantified by $\alpha_0$, increases in course of
renormalization. The renormalization process should be stopped when the
renormalized value of $\alpha_0$ becomes of order of unity, so that the
perturbative expansion of the beta
function fails. The corresponding scale is the zero-energy mean free path,
which we denote $l_0$. To find this length, we express $\Lambda$ as function of
$\alpha_0$ in Eq.\ (\ref{RGalpha0}) and integrate from the initial value of
$\alpha_0$ to $1$. This yields
\begin{equation}
 l_0
  = a \sqrt{\alpha_0} e^{1/2 \alpha_0}.
 \label{l0}
\end{equation}
The universality of the two-loop equation is evident from Eq.\ (\ref{l0}). The
first loop
contribution determines the exponential factor in $l_0$ while the second loop
gives $\sqrt{\alpha_0}$ in the pre-exponent. This parametric dependence of $l_0$ can not
depend on the regularization scheme. On the other hand, the third loop would fix the
numerical prefactor in Eq.\ (\ref{l0}).
However, the value of the ultraviolet length $a$ is itself defined only up to a number
within the framework of the linearized Dirac Hamiltonian model.

At scales shorter than the mean free path $l_0$, the renormalized value of
$\alpha_0$ is given by
\begin{equation}
 \alpha_0(\Lambda)
  = \frac{1}{2 \ln(l_0/\Lambda) + \ln \ln (l_0/\Lambda)}.
 \label{alpha0renorm}
\end{equation}
As long as $L\ll l_0$ [and thus $\alpha_0(L)\ll 1$],
we can describe the transport properties by the distribution function
(\ref{PTUB}) with the
renormalized value $\alpha_0(L)$ and $\epsilon = 0$. It is worth noting that the
lowest order perturbation theory used for derivation of Eq.\ (\ref{PTUB})
in combination with the RG result (\ref{alpha0renorm}) provides the best possible
accuracy within the framework of disordered Dirac Hamiltonian.
Specifically, the second-order terms in the perturbative expansion of $P(T)$ in powers of
$\alpha_0(L)$ would generate the contribution of the same order
as that of the third-loop correction to Eq.\ (\ref{RGalpha0}).
The latter, however, depends on the regularization scheme and hence
is nonuniversal, as discussed above.

A small but non-zero energy does not change the qualitative behavior of the
system, as long as the RG flow is terminated by the system size.
We refer to this situation as ``ultraballistic regime''.
The energy gets renormalized according to Eq.\ (\ref{RGe}), which is
universal only in the one-loop order (see Appendix \ref{App:twoloop}). Using the
result (\ref{alpha0renorm}), we solve the RG equation for energy and obtain
\begin{equation}
 \epsilon(\Lambda)
  = \frac{\epsilon}
    {\sqrt{\alpha_0[2 \ln(l_0/\Lambda) + \ln \ln (l_0/\Lambda)]}}.
 \label{erenorm}
\end{equation}
It is worth mentioning that the renormalized coupling (\ref{alpha0renorm})
and the renormalized energy (\ref{erenorm}) are related via
\begin{equation}
 \label{e-alpha-renorm}
 \frac{\epsilon^2(\Lambda)}{\epsilon^2}
  = \frac{\alpha_0(\Lambda)}{\alpha_0}.
\end{equation}
The value $\epsilon(L)$ is to be substituted into Eq.\ (\ref{PTUB}) along with
the renormalized value of $\alpha_0(L)$. This yields the full description of
transport properties for the system in the ultraballistic regime. In particular,
the conductance and the Fano factor are
\begin{align}
 G
  &=\frac{4 e^2}{\pi h}\; \frac{W}{L} \left[
      1 + \frac{2\alpha_0
      +c_1 (\epsilon L)^2}{\alpha_0[2 \ln(l_0/L) + \ln \ln (l_0/L)]}
    \right],
 \label{GUB} \\
 F
  &=\frac{1}{3} \left[
      1 + \frac{c_2 (\epsilon L)^2}{\alpha_0[2 \ln(l_0/L) + \ln \ln (l_0/L)]}
    \right],
 \label{FUB}
\end{align}
with the constants $c_{1,2}$ given by Eqs.\ (\ref{c1}) and (\ref{c2}).

When the initial (bare) value of energy is increased,
the renormalized energy eventually
becomes comparable to the effective bandwidth $1/\Lambda$
before the running scale $\Lambda$ reaches $L$ (and still before the
disorder coupling $\alpha_0(\Lambda)$ reaches unity).
The length scale at which
$\epsilon(\Lambda) = 1/\Lambda$ plays the role of the effective Fermi
wavelength $\lambda$. (Indeed, in the absence of disorder, energy is not renormalized
and $\lambda = 1/\epsilon$.)
Using Eq.\ (\ref{erenorm}), we find
\begin{equation}
 \lambda
  = \frac{1}{\epsilon} \sqrt{2 \alpha_0 \ln(\epsilon/\gamma)},
 \label{lambda}
\end{equation}
where $\gamma$ is the characteristic disorder-induced energy scale
\begin{equation}
 \gamma
  = \sqrt{\alpha_0}/l_0
  = \Delta e^{-1/2\alpha_0},
 \label{gamma}
\end{equation}
$\Delta = 1/a$ is the initial bandwidth of the model, and we assumed that
$\epsilon \gg \gamma$. For $\epsilon \alt \gamma$, the role of the
wavelength is played by the mean free path $l_0$.  Note that Eqs.\
(\ref{lambda}) and (\ref{gamma}) have the same two-loop accuracy as Eqs.\
(\ref{l0}) -- (\ref{erenorm}); in particular, the absence of a double logarithm
term in Eq.\ (\ref{lambda}) and of $\alpha_0$ in the pre-exponent in Eq.\
(\ref{gamma}) is fully controllable. According to Eqs.\ (\ref{lambda}),
(\ref{e-alpha-renorm}), the renormalized values of the coupling constant and the
energy at the scale of the wave length are given by
\begin{equation}
 \label{e-lambda-a0-lambda}
 \epsilon(\lambda)
  = \frac{\epsilon}{\sqrt{2\alpha_0\ln(\epsilon/\gamma)}},
 \qquad
 \alpha_0(\lambda)
  = \frac{1}{2\ln(\epsilon/\gamma)}.
\end{equation}

\begin{figure}
 \centerline{\includegraphics[width=0.95\columnwidth]{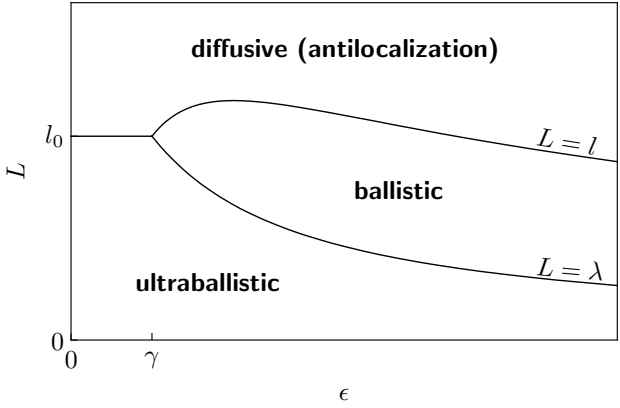}}
 \caption{Schematic ``phase diagram'' of various transport regimes in the
graphene sample with random scalar potential. The lines indicate crossovers
between corresponding regimes. The shortest sample exhibits ultraballistic
transport with the conductance and Fano factor given by Eqs.\
(\protect\ref{GUB}) and (\protect\ref{FUB}) respectively. When the length of the
sample exceeds Fermi wave length (\protect\ref{lambda}), ballistic results
(\protect\ref{GB}) and (\protect\ref{FB}) apply. In a sample longer than the
mean free path (\protect\ref{MFP}), diffusive regime establishes with the Drude
conductivity (\protect\ref{Drude}) and the Dorokhov distribution of transmission
eigenvalues (\protect\ref{Dorokhov}). The conductivity experiences symplectic
antilocalization in this case.}
 \label{Fig:phase}
\end{figure}

In Fig.\ \ref{Fig:phase} we show the phase
diagram of various transport regimes.
If $\epsilon \lesssim \gamma$ and $L\ll l_0 $ or, alternatively,
$\epsilon \gtrsim \gamma$ and $L\ll \lambda$, the renormalization terminates
by the system size, $\Lambda=L$, and the system is in the ultraballistic regime
discussed above [see Eqs. (\ref{GUB}) and (\ref{FUB})].
If $\epsilon \gg \gamma$ and $\lambda \ll L \ll l$, the renormalization stops
at $\Lambda = \lambda$ and the running scale does not reach $L$.
We refer to this case as ``ballistic regime'', since the system size is still smaller
than the mean free path $l$,
\begin{equation}
 l
  = \frac{\lambda}{\pi \alpha_0(\lambda)}
  = \frac{\sqrt{\alpha_0}}{\pi \epsilon} [2 \ln (\epsilon/\gamma)]^{3/2}.
 \label{MFP}
\end{equation}
This value\cite{OurPRB} of the mean free path corresponds to the imaginary
part of the
electron self-energy calculated in the Born approximation with renormalized
coupling constant $\alpha_0(\lambda)$. Note that for the
model with random scalar potential, the transport mean free path, which
determines diffusion coefficient, is twice longer, $l_{\mathrm{tr}} = 2 l$.

In the ballistic regime, the
renormalized energy is such that $\epsilon(\lambda) L = L/\lambda \gg 1$. This
means that we have to use the high-energy results of Sec.\ \ref{high-ballistic}.
In particular,
with the renormalized parameters, the conductance and the Fano factor, Eqs.\
(\ref{Gelarge}) and (\ref{Felarge}), become
\begin{align}
 G
  &= \frac{e^2}{h}\; \frac{W}{\lambda} \left[
      1 + \frac{\sin (2 L/\lambda - \pi/4)}{2\sqrt{\pi} (L/\lambda)^{3/2}}
      -\frac{L}{4 l}
    \right],
 \label{GB}\\
 F
  &= \frac{1}{8} \left[
      1 - \frac{9\sin (2 L/\lambda - \pi/4)}{2\sqrt{\pi} (L/\lambda)^{3/2}}
      +\frac{3 L}{4 l}
    \right].
 \label{FB}
\end{align}
In the expressions (\ref{GB}) and (\ref{FB}) there are two corrections to the
leading term. The first (oscillating) correction exists in the clean limit and
is small provided $L \gg \lambda$. The second correction due to disorder is
small only if $L \ll l$. This imposes the natural upper bound on the ballistic
regime: if the system size exceeds the mean free path, electron transport
becomes diffusive. In this case, the system is naturally characterized by
the conductivity $\sigma$, which determines the conductance via the Ohm's law,
$G=\sigma W/L$. The Drude expression for the conductivity reads \cite{Aleiner06}
\begin{equation}
 \sigma
  = \frac{4 e^2}{\pi h \alpha_0(\lambda)}
  = \frac{8 e^2}{\pi h }\, \ln (\epsilon/\gamma).
 \label{Drude}
\end{equation}
The distribution function of transmission eigenvalues in the diffusive regime is
the same as in a usual quasi-one-dimensional metallic sample \cite{Dorokhov83}
\begin{equation}
 P(T)
  = \frac{W}{2\pi L}\, \frac{g}{T \sqrt{1 - T}},
 \label{Dorokhov}
\end{equation}
with the dimensionless conductivity $g = (\pi h / 4 e^2) \sigma$. Taking into
account interference effects leads to $L$ dependence of $g$ in this formula, as
we are going to discuss.

\subsection{Single parameter scaling for random potential at zero energy}
\label{Sec:single-parameter-scaling}

\begin{figure}
 \centerline{\includegraphics[width=0.95\columnwidth]{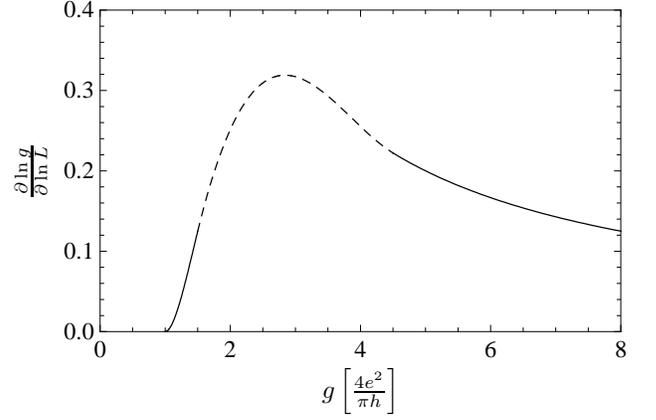}}
 \caption{Unified scaling function (\protect\ref{scaling-g})
for both ultraballistic and diffusive regimes at zero energy in the case of the
random potential disorder.}
 \label{Fig:scaling}
\end{figure}

Remarkably, the transmission distribution function at zero energy appears to be
the same in
ultraballistic and diffusive limits. In both cases it has the form of Dorokhov
distribution (\ref{Dorokhov}) with the parameter
\begin{equation}
 g
  = \begin{cases}
      1 + 2 \alpha_0(L), & \text{ultraballistic}, \\
      \dfrac{\pi h}{4 e^2}\; \sigma(L), & \text{diffusive},
    \end{cases}
\end{equation}
which has the meaning of the dimensionless conductivity. In the ultraballistic
regime, the scaling of $g$ is induced via renormalization of $\alpha_0$
according to equation (\ref{RGalpha0}) while in the diffusive limit, $g\gg 1$
acquires antilocalization corrections characteristic for a disordered system of
symplectic symmetry. This allows us to infer a unified scaling law covering both
limiting cases
\begin{equation}
 \frac{\partial \ln g}{\partial \ln L}
  = \begin{cases}
      (g - 1)^2-\dfrac{1}{2}(g - 1)^3, & g-1 \ll 1, \\
      \dfrac{1}{g}, & g \gg 1.
    \end{cases}
 \label{scaling-g}
\end{equation}
The scaling function (\ref{scaling-g}) is depicted in Fig.\ \ref{Fig:scaling}.
It is qualitatively similar to the numerical results of Ref.\
\onlinecite{Bardarson07}.

The applicability of the Dorokhov distribution to the diffusive system in the
considered geometry requires a comment. The original derivation given by
Dorokhov \cite{Dorokhov83} assumes a diffusive system of a quasi-1D geometry
(a thick wire), with $W \ll L$. On the other hand, our geometry is entirely
different, $W \gg L$. This difference is, however, of minor importance for the
statistics of charge transfer as long as the system is a good metal. Indeed,
there exist alternative derivations of the Dorokhov statistics that are based on
the semiclassical Green function formalism \cite{Nazarov94, Nazarov99}, on the
sigma-model approach \cite{Gutman03} or on the kinetic theory of fluctuations
\cite{Gutman04} and do not require any assumption concerning the aspect ratio
of the sample.

We stress that the ultraballistic asymptotics of the beta function in Eq.\
(\ref{scaling-g}) is only valid for Gaussian white-noise statistics of random
potential. The interpolation between the two asymptotics of the beta function in
Fig.\ \ref{Fig:scaling} implicitly assumes a smooth crossover between the two
regimes (in particular, without any intermediate fixed points), as suggested by
numerical simulations. \cite{Bardarson07, Koshino07, San-Jose07}

The scaling function (\ref{scaling-g}) characterizes the evolution of the
dimensionless conductivity with increasing $L$.  Whether the full distribution
function $P(T)$ retains exactly its form (\ref{Dorokhov}) (parameterized by $g$
only) in the crossover remains an open question. Strictly speaking, what we know
at the moment is that this form of $P(T)$ emerges (i) in the clean limit, (ii)
in the ultraballistic regime within the first order in $\alpha_0(L)$, and (iii)
in the diffusive regime. On this basis, one could speculate that this might be
an exact statement for the whole crossover. Clearly, this is only a hypothesis
that requires further verification.

\subsection{Random vector potential}
\label{Sec:vector-potential}

Let us now consider the situation when the only disorder in the sample is
the random vector potential (characterized by the couplings $\alpha_x$ and
$\alpha_y$). This situation is physically realized when disorder
is due to random corrugations of the graphene sheet (ripples).
The one-loop RG equations for random vector potential read\cite{Ludwig}
\begin{align}
&\frac{\partial \alpha_x}{\partial \ln \Lambda} =
\frac{\partial \alpha_y}{\partial \ln \Lambda}=0, \\
&\frac{\partial \epsilon}{\partial \ln \Lambda} =
\epsilon \left(\alpha_x +\alpha_y\right).
\end{align}
In fact, the beta function for the disorder couplings is identically
zero in all loops, \cite{Ludwig,Guruswamy}
i.e., the random vector potential is not renormalized.
Since the couplings do not change with growing system size, the energy follows a power-law:
\begin{equation}
 \epsilon(\Lambda)= \epsilon\ \left(\frac{\Lambda}{a}\right)^{\alpha_\perp},
\end{equation}
where $\alpha_\perp:=\alpha_x+\alpha_y$.
For not too high energies, the RG flow terminates by the system size $L$
(ultraballistic regime), so that
\begin{equation}
 \epsilon(L)L= \epsilon L (L/a)^{\alpha_\perp} \ll 1.
\label{RVPEL}
\end{equation}

As demonstrated in Sec.\ \ref{Sec:first-correction}, at zero energy the
lowest-order perturbative correction to the transport coefficients is absent in
the case when the only disorder is vector potential. Now we present a general
argument showing that any given configuration of the vector potential
$\mathbf{A}(x,y)$ does not affect transport properties of the system at zero
energy.

The zero-energy Dirac equation takes the following form in the presence of
vector potential,
\begin{equation}
 \bm{\sigma}(\mathbf{p} - \mathbf{A}) \Psi
  = 0.
\end{equation}
We fix the gauge by requiring that $\nabla \mathbf{A} = 0$ in the bulk of the
sample and normal component of $\mathbf{A}$ vanishes at the boundary of the
sample. This gauge is widely used in the theory of superconductivity and is
referred to as the London gauge in that context. In this particular gauge we can
express vector potential using a scalar function $\phi(x,y)$ as
\begin{equation}
 A_x
  = \frac{\partial \phi}{\partial y},
 \qquad
 A_y
  = -\frac{\partial\phi}{\partial x}.
\end{equation}
The boundary conditions allow us to fix $\phi = 0$ at the edges of the sample.
The function $\phi$ is related to the magnetic field $B = \partial_x A_y -
\partial_y A_x$ by the Poisson equation $\nabla^2 \phi = -B$. The existence of a
solution to such an equation follows from an equivalent electrostatics problem:
finding the potential of the charge distribution with a given density inside a
metallic cavity.

Now we do a pseudo-gauge transformation introducing the new wave function $\bar
\Psi$ according to $\Psi = e^{\sigma_z \phi} \bar \Psi$. For this new function,
the Dirac equation becomes
\begin{equation}
 e^{-\sigma_z \phi} \bm{\sigma} \mathbf{p} \bar\Psi
  = 0,
\end{equation}
which is equivalent to the free Dirac equation with no magnetic field. The
boundary conditions of the London gauge fix $\phi = 0$ outside of the sample.
Thus $\Psi = \bar\Psi$ in the leads. The transfer matrix of the whole system,
and hence all the transport properties, is not influenced by the vector
potential. This result was obtained in Ref.\ \onlinecite{Fogler} for a
particular configuration of vector potential. Recently, we have become aware of
an alternative proof of the general statement by Titov. \cite{Titov-private}
The immunity of the transport properties to the vector potential holds despite
the fact that the random vector potential problem represents a critical theory 
\cite{Ludwig} with the multifractal wave function $\Psi(x,y)$ and a spectrum of
multifractal exponents governed by the disorder strength $\alpha_\perp$ (for
review see Ref.\ \cite{MirlinRMP}).

We will use the results of Sec.\ \ref{Sec:clean} obtained for the clean sample.
Disorder, however, affects the transport properties through the
disorder-dependent renormalization of energy. Thus, in the ultraballistic
regime, we can use equations (\ref{PTe}) and (\ref{GFsmallE}) of Sec.\
\ref{Sec:CleanLowEnergies} with $\epsilon L$ replaced by its renormalized value
given by Eq.\ (\ref{RVPEL}).

When the system size $L$ is larger than the Fermi wavelength $\lambda$, the
renormalization of energy stops by the bandwidth. The value of $\lambda $ is
found from the equation $1/\lambda=\epsilon(\lambda)$, yielding
\begin{equation}
 \lambda
  = \frac{1}{\epsilon} \left(
      \frac{\epsilon}{\Delta}
    \right)^{\alpha_\perp/(1+\alpha_\perp)}.
 \label{RVPlambda}
\end{equation}
The $L$-independent renormalized energy $\epsilon(\lambda)$ is thus given by
$
\epsilon(\lambda)=\epsilon \left(\Delta/\epsilon\right)^{\alpha_\perp/(1+\alpha_\perp)}.
$

To calculate the transport coefficients in the ballistic regime, we substitute the
product $\epsilon(\lambda)L=L/\lambda \gg 1$ along with the couplings $\alpha_x$ and $\alpha_y$
into Eqs.\ (\ref{Gelarge}) and (\ref{Felarge}),
which yields
\begin{align}
 G
  &= \frac{e^2}{h}\; \frac{W}{\lambda} \left[
      1 + \frac{\sin (2 L/\lambda - \pi/4)}{2\sqrt{\pi} (L/\lambda)^{3/2}}
      - \frac{\pi L}{4\lambda}(\alpha_x + 3\alpha_y)
          \right],
 \label{GBRVP}\\
 F
  &= \frac{1}{8} \left[
      1 - \frac{9\sin (2 L/\lambda - \pi/4)}{2\sqrt{\pi} (L/\lambda)^{3/2}}
      +\frac{\pi L}{4\lambda} (
       3\alpha_x + 13\alpha_y
      )
    \right].
 \label{FBRVP}
\end{align}

Using  Eq.\ (\ref{RVPlambda}) we find the mean free path
\begin{equation}
 l(\epsilon)
  = \frac{\lambda}{\pi\alpha_\perp}
  = \frac{1}{\pi\alpha_\perp \epsilon} \left(
      \frac{\epsilon}{\Delta}
    \right)^{\alpha_\perp/(1+\alpha_\perp)}.
\label{RVPl}
\end{equation}
The system becomes diffusive when $L\gtrsim l$. In contrast to the case of
scalar potential discussed in Sec.\ \ref{RSP}, for random vector potential there
is no direct crossover between the ultraballistic and diffusive regimes (at zero
energy the mean free path diverges).

\begin{figure}
 \centerline{\includegraphics[width=0.95\columnwidth]{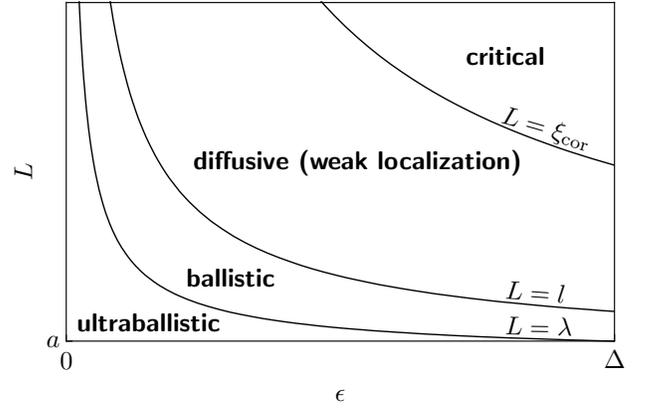}}
 \caption{Schematic ``phase diagram'' for the sample with random vector
potential. Ultraballistic and ballistic regimes are similar to the case of
random scalar potential. However, the time-reversal symmetry is broken and the
system exhibits weak (second-loop) localization in the diffusive regime, which
eventually drives it to the critical state characteristic for the quantum Hall
transition. There is no disorder-induced energy scale in the case of random
vector potential, thus we plot the phase diagram for all energies up to the
band width $\Delta$, which is exponentially larger than $\gamma$ in Fig.\
\protect\ref{Fig:phase}. In the case of random mass, the phase diagram is
qualitatively the same.}
 \label{phaseRVPM}
\end{figure}

At finite energy the system belongs to the unitary symmetry class. The
corresponding field-theory possesses a topological term,\cite{OurPRL} which
drives the system to the quantum-Hall critical point with the universal value of
conductivity $\sigma = (0.5 \div 0.6)\, 4 e^2/h$. Since the disorder coupling
$\alpha_\perp$ stays non-renormalized, the Drude conductivity\cite{OurPRB} is
given by the
Born approximation with the bare value of $\alpha_\perp$,
\begin{equation}
 \sigma
  = \frac{2 e^2}{\pi h \alpha_\perp},
 \label{DrudeRVP}
\end{equation}
and does not depend on energy. The criticality is achieved only at very large
scales $L \sim \xi_{\text{cor}}$. The quantum-Hall correlation length
$\xi_{\text{cor}}$ is of the order of the unitary-class
(second-loop) localization length
$\xi \propto \exp(g^2)$ at which all states would be
localized in the absence of the topological term,
\begin{equation}
 \xi_{\text{cor}}
  \sim l(\epsilon)\; e^{1/4\alpha_\perp^{2}}.
\label{xilocU}
\end{equation}
The phase diagram for the random vector potential (Fig.\ \ref{phaseRVPM})
contains four regions: ultraballistic ($0<L<\lambda$), ballistic
($\lambda<L<l$), diffusive ($l<L<\xi_{\text{cor}}$), and critical
($L>\xi_{\text{cor}}$).

\subsection{Random mass}
\label{Sec:random-mass}

Let us now discuss the transport properties in a situation when disorder is
modeled solely by the random mass term ($\alpha_z$ coupling). We are not aware
of physical realizations of such disorder in graphene. Nevertheless, we will
consider this case for the sake of completeness.

The RG equations for the random mass read
\begin{equation}
 \frac{\partial \alpha_z}{\partial \ln \Lambda}
  = -2\alpha_z^2,
 \qquad
 \frac{\partial \epsilon}{\partial \ln \Lambda}
  = \epsilon \alpha_z,
\end{equation}
It is worth mentioning that there is an exact (valid in all loops)
relation\cite{Ludwig}
between the beta function $\beta_z$ for the random-mass coupling $\alpha_z$
and $\beta_0$ for the random potential $\alpha_0$:
\begin{equation}
 \beta_z (\alpha_z) = -\beta_0(-\alpha_z).
\end{equation}
However, we do not need the two-loop result for the random mass, since
$\alpha_z(\Lambda)$ decreases with growing $\Lambda$ and therefore the
second-loop term never becomes important.

The one-loop RG equations for $\alpha_z$ and $\epsilon$ are solved by
\begin{align}
 \alpha_z(\Lambda)
  &= \frac{\alpha_z}{1+2\alpha_z \ln(\Lambda/a)}, \label{alphazL}\\
 \epsilon(\Lambda)
  &= \epsilon \sqrt{\frac{\alpha_z}{\alpha_z(\Lambda)}}
  = \epsilon \sqrt{1+2\alpha_z \ln \frac{\Lambda}{a}}.
 \label{zepsilonL}
\end{align}
Thus, while $\alpha_z(\Lambda)$ decreases with growing length scale $\Lambda$,
the energy $\epsilon(\Lambda)$ becomes larger. Energy reaches the bandwidth
at the Fermi wave length scale $\lambda$ determined by $\epsilon(\lambda) =
1/\lambda$. This yields
\begin{align}
 \lambda
  &= \frac{1}{\epsilon} \frac{1}{\sqrt{1+2\alpha_z \ln (\Delta/\epsilon)}},
 \label{zlambda}\\
 \epsilon(\lambda)
  &= \epsilon \sqrt{1+2\alpha_z \ln (\Delta/\epsilon)}. \label{zepsilonlambda}
\end{align}
At this scale the disorder coupling becomes
\begin{equation}
 \alpha_z(\lambda) =\frac{\alpha_z}{1+2\alpha_z \ln \Delta/\epsilon}
\label{alphazlambda}
\end{equation}

To describe the transport properties in the ultraballistic regime ($L \ll
\lambda$), we substitute Eqs.\ (\ref{alphazL}) and (\ref{zepsilonL}) taken at
$\Lambda = L$ into Eq.\ (\ref{PTUB}) and find
\begin{align}
 G
  &=\frac{4 e^2}{\pi h}\; \frac{W}{L} \Big\{
      1 - \frac{2\alpha_z}{1+2\alpha_z \ln(L/a)}\notag \\
      &+ c_1 (\epsilon L)^2[1+2\alpha_z \ln(L/a)]
    \Big\},
 \label{RMGUB} \\
 F
  &=\frac{1}{3} \left\{
      1 +c_2 (\epsilon L)^2[1+2\alpha_z \ln(L/a)]
    \right\},
 \label{RMFUB}
\end{align}

When the system size exceeds the wavelength, $L\gg\lambda$, we use
the values for $\alpha_z$ and $\epsilon$
given by Eqs.\ (\ref{alphazlambda}) and (\ref{zepsilonlambda}).
The system becomes diffusive when
\begin{equation}
 L\gtrsim l= \frac{\lambda}{\pi\alpha_z(\lambda)}=
\frac{1}{\pi\alpha_z \epsilon} \sqrt{1+2 \alpha_z \ln \frac{\Delta}{\epsilon}}
\end{equation}
In the ballistic regime $\lambda\ll L\ll l$ we get
\begin{align}
 G
  &= \frac{e^2}{h}\; \frac{W}{\lambda} \left[
      1 + \frac{\sin (2 L/\lambda - \pi/4)}{2\sqrt{\pi} (L/\lambda)^{3/2}}
      - \frac{3L}{4l}
          \right],
 \label{GBRM}\\
 F
  &= \frac{1}{8} \left[
      1 - \frac{9\sin (2 L/\lambda - \pi/4)}{2\sqrt{\pi} (L/\lambda)^{3/2}}
      + \frac{13L}{4l}
    \right],
 \label{FBRM}
\end{align}
where $\lambda$ is given by Eq. (\ref{zlambda}).

At zero energy, the system belongs to the superconducting symmetry class D (see,
e.g., Ref.\ \onlinecite{Bocquet}). Finite energy drives the system to the
unitary symmetry class A, similarly to the case of random vector potential.
Again, weak (second-loop) localization leads to the quantum-Hall
critical point at very large scales  $L\gtrsim \xi_{\text{cor}}$. The Drude
conductivity \cite{OurPRB} is given by the Born
approximation with the renormalized value of $\alpha_z$ from Eq.\
(\ref{alphazlambda}),
\begin{equation}
 \sigma
  = \frac{4 e^2}{3 \pi h} \left[
      \frac{1}{\alpha_z} + 2\ln\frac{\Delta}{\epsilon}
    \right].
 \label{zDrude}
\end{equation}
The corresponding quantum-Hall correlation length is then given by
\begin{equation}
 \xi_{\text{cor}}
  \sim l(\epsilon) \exp\left[\frac{1}{9} \left(
      \frac{1}{\alpha_z} + 2\ln\frac{\Delta}{\epsilon}
    \right)^{2}\right].
 \label{zxilocU}
\end{equation}
Thus the case of random mass disorder is very similar to the case of random
vector potential. We have four regimes: ultraballistic ($0<L<\lambda$),
ballistic ($\lambda<L<l$), diffusive ($l<L<\xi_{\text{cor}}$), and critical
($L>\xi_{\text{cor}}$). The schematic phase diagram is the same as for the
random vector potential case presented in Fig.\ \ref{phaseRVPM}. There is no
direct crossover between the ultraballistic and diffusive regimes. This is
related to the fact that at zero energy the system is ultraballistic for
arbitrary length $L$.

\subsection{Generic disorder}
\label{Sec:generic-disorder}

Finally, let us consider the case of generic disorder, when all disorder
couplings are present. In fact, even if only two of the three coupling constants
$\alpha_0$, $\alpha_\perp$, and $\alpha_z$ are present at the initial
ultraviolet scale $a$, the third one always becomes non-zero with growing system
size, \cite{Ludwig} see Eqs. (\ref{RGa0}), (\ref{RGaperp}), and (\ref{RGaz}).
The system belongs to the unitary symmetry class at all energies and
falls into the quantum Hall universality class. \cite{Ludwig, OurPRL}
Physically, this situation is realized in graphene when, e.g., both long-range
vector (ripples) and scalar (charged impurities) potential are present.
\cite{OurPRL, OurEPJ, OurQHE}

As discussed in Appendix \ref{App:twoloop}, when two or more coupling constants
are nonzero, the second-loop beta-function becomes non-universal. Therefore, we
will deal here with the one-loop RG equations. The solution of the set of
coupled RG equations (\ref{RGa0}), (\ref{RGaperp}), and (\ref{RGaz}) is analyzed
in Appendix \ref{App:oneloop}. It turns out that when the initial values of the
couplings are of the same order, after renormalization the coupling $\alpha_0$
(corresponding to the scalar potential) dominates. In particular, at zero energy
the renormalization stops at the scale $l_0$ when $\alpha_0(l_0) \sim 1$, while
the other two couplings are still much smaller than unity (suppressed by a
logarithmic factor), $\alpha_\perp(l_0), \alpha_z(l_0) \simeq (9/8)
|\ln\alpha_0|^{-1}$, where $\alpha_0 \ll 1$ is the bare value of the couplings.

\begin{figure}
 \centerline{\includegraphics[width=0.95\columnwidth]{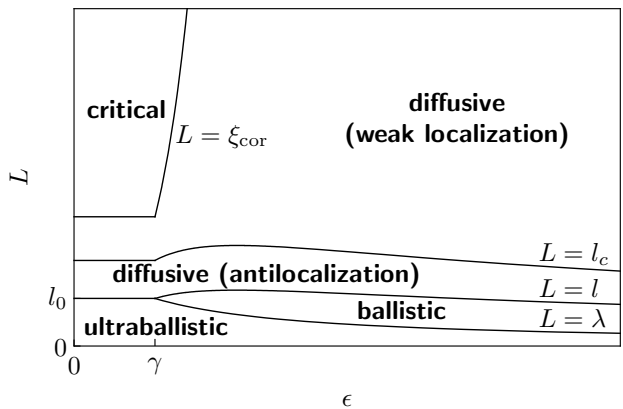}}
 \caption{Schematic ``phase diagram'' for the case when more than one disorder
type is present in the system. Random scalar potential $\alpha_0$ becomes
dominant in the course of ballistic renormalization (see Appendix
\protect\ref{App:oneloop}); therefore, the ultraballistic, ballistic, and
lowest part of the diffusive regime are similar to the diagram Fig.\
\protect\ref{Fig:phase}. Once the diffusion is established, the antilocalization
starts but it proceeds only till the length $l_c$ at which the time-reversal
symmetry breaks down.\protect\cite{McCann} At longer scales the system falls
into the unitary symmetry class and exhibits weak (second-loop) localization. At
exponentially long scale $\xi_\text{cor}$ the quantum Hall critical state
establishes.}
 \label{phase-generic}
\end{figure}

The phase diagram for the case of generic disorder (Fig.\ \ref{phase-generic})
contains four regimes, similarly to the cases of random vector potential and
random mass: in addition to the ultraballistic, ballistic, and diffusive
regimes, there is a regime of quantum-Hall criticality. On the other hand, at
variance with the random vector potential and random mass problems, the
diffusive regime consists of two subregimes (with weak antilocalization and weak
localization corrections to the Drude conductivity, respectively). Indeed, as
discussed above, at the border of the diffusive regime ($L \sim l$) the dominant
coupling is $\alpha_0$, which corresponds to the symplectic symmetry
class. \cite{footnote-noantilocalization}
At larger scales ($L > l_c \sim l_0 |\ln \alpha_0|^{-1/2}$ for $\epsilon = 0$),
the gap in the Cooperon modes due to the couplings $\alpha_\perp$ and $\alpha_z$
becomes important \cite{McCann} and only diffuson modes remain, restoring the
unitary symmetry and leading to the second-loop localizing correction to the
conductivity. When the renormalized conductivity drops down to the value of the
order $e^2/h$, the critical quantum Hall regime sets in. It is also worth
mentioning that, similarly to the case of random scalar potential, at lowest
energies (including $\epsilon=0$) the ultraballistic regime crosses over
directly into the diffusive/critical regime.

\subsection{Additional comments}
\label{Sec:additional_comments}

Throughout this paper, we have considered a somewhat idealized theoretical
model. Specifically, we have neglected (i) intervalley scattering,
 (ii) momentum dependence of scattering amplitude characteristic for scatterers
with $1/r$ potentials (like charged impurities or ripples), and (iii)
electron-electron interaction. We are now going to discuss, on the qualitative
level, a possible influence of these effects on our results.

{\it (i) Intervalley scattering.}
As discussed in the beginning of the paper, the fact that the dominant
disorder scattering in experimentally studied graphene samples is of
intra-valley nature follows directly from the observed anomalous, odd-integer
quantum Hall effect. The dominance of the intra-valley scattering also explains
why the localization is not observed at the Dirac point down to very low
temperatures. Therefore, the model of decoupled valleys considered in this work
is not only of theoretical interest but is also directly relevant to
experiments.

Still, in any realistic system some amount of inter-valley scattering will be
present, so that it is natural to ask what its influence will be. The weakness
of the inter-valley scattering implies that the corresponding mean free path
$l_{\rm inter}$ is much larger than the mean free path $l$ (induced by the
intra-valley scattering and considered in the paper). This means that the
scale $l_{\rm inter}$ is generically located far in the diffusive (or critical)
regime. The results in the ultraballistic and ballistic regimes, as well as in
a parametrically broad window in the diffusive and critical regimes, remain
essentially unaffected by the inter-valley scattering. At very large
distances, $L\gg l_{\rm inter}$, the intravalley scattering will strongly affect
the behavior, generically inducing the localization (except for a special case
of chiral disorder at the Dirac point\cite{OurPRB}).

{\it (ii) $1/r$ impurities.}  
Most of realistic candidates for long-range scatterers in graphene
samples, such as charged impurities and ripples, are characterized by $1/r$
potentials. As was shown in Ref.\ \onlinecite{OurPRB}, there is no ballistic RG 
for this type of scatterers; the scale- and energy-dependences of
disorder-induced effects in the ballistic regime are governed simply by the
energy dependence of the cross-section of an individual scatterer. 
With these modifications, all the considerations in our paper remain
applicable. In particular, all the phase diagram remain qualitative unchanged;
one should just use the appropriate values of the wave length $\lambda$ and
of the mean free path $l$.

{\it (iii) Electron-electron interaction.}
The effect of el\-ect\-ron-el\-ect\-ron interaction on the system of disordered
Dirac fermions constitutes in general a very complex problem. In the clean case,
the interaction induces a logarithmic correction to the velocity\cite{Gonzalez}
that can be treated within an RG scheme similar to the ballistic disorder RG
used in this work. In the disordered case, a unified ballistic RG emerges
\cite{Foster06, Foster08} describing renormalization of disorder couplings
and of the interaction. In Ref.\ \onlinecite{Foster08} corresponding one-loop RG
equations are derived for time-reversal-invariant disorder
and in the limit of large number of valleys (simplifying the theoretical
treatment). One can use this interaction-modified RG values for renormalized
couplings entering our results in the ultraballistic and ballistic regimes;
this analysis is, however, beyond the scope of the present work.

It should be stressed that, as discussed above, the ballistic RG is
redundant for scatterers with $1/r$ potentials, like charged impurities or
ripples. In this situation, inclusion of interaction does not lead to any
essential modifications as long as the system is in the ultraballistic or
ballistic regime.

\section{Summary}
\label{Sec:summary}

In this paper, we have analyzed transport properties of a graphene sample
in the ``wide and short'' geometry, $W \gg L$, with disorder effects restricted
to intra-valley scattering. Starting from the clean limit and using the
transfer-matrix technique, we have analyzed the evolution of the transmission
distribution $P(T)$ and, in particular, of the conductance $G$ and the Fano
factor $F$, with increasing system size $L$. To take the randomness into
account, we have developed a perturbative treatment of the transfer-matrix
equations supplemented by an RG formalism describing the renormalization of
disorder couplings. This has allowed us to get complete analytical description
of the transport properties of graphene in the ultraballistic ($L\ll \lambda$)
and ballistic ($\lambda \ll L \ll l$) regimes.
We have also constructed ``phase diagrams'' of different transport regimes
(ultraballistic, ballistic, diffusive, critical) for graphene with various types
(symmetries) of intra-valley disorder.

\begin{acknowledgments}
We are grateful to M.~Titov, A.W.W.~Ludwig, P.~San-Jose, E.~Prada, and
H.~Schomerus for valuable discussions. The
work was supported by the Center for Functional Nanostructures of the Deutsche
Forschungsgemeinschaft. The work of I.V.G. was supported by the EUROHORCS/ESF
EURYI Awards scheme. I.V.G. and A.D.M.  acknowledge kind hospitality of the
Isaac Newton Institute for Mathematical Sciences of the Cambridge University
during the program ``Mathematics and physics of Anderson localization: 50 years
after'', where a part of this work was performed.
\end{acknowledgments}

\appendix

\section{Oscillations at high energies}
\label{App:oscillations}

In this Appendix we consider transport properties of a clean graphene sample in
the limit of high energies, $\epsilon L \gg 1$. We will find the next term in
the inverse energy expansion of the generating function $\mathcal{F}(z)$ and
the distribution function $P(T)$, which yields an oscillatory correction to the
results (\ref{Fz}) and (\ref{PTelarge}). Our starting point is the exact
expression for the generating function Eq.\ (\ref{F0int}) that we rewrite using
the parameterization (\ref{u-index}),
\begin{equation}
 \mathcal{F}(z)
  = \frac{W \epsilon}{\pi} \int_0^1 \frac{u\, du}{\sqrt{1-u^2}} \left[
      \cos^2(u \epsilon L) + \frac{\sin^2(u \epsilon L)}{u^2} - z
    \right]^{-1}.
\end{equation}
Trigonometric functions in the integrand rapidly oscillate. To take advantage
of this property, we represent the integrand as a sum over Fourier harmonics
$\cos(n u\epsilon L)$. The first and the second terms of such Fourier
expansion are
\begin{multline}
 \mathcal{F}(z)
  = \frac{W \epsilon}{\pi \sqrt{1-z}}
    \int_0^1 \frac{u^2\, du}{\sqrt{(1 - u^2)(1- z u^2)}} \Biggl[
      1 \\
      +\frac{2 \bigl(u \sqrt{1 - z} - \sqrt{1-z u^2} \bigr)^2}{1 - u^2}
        \cos(2 u \epsilon L)
    \Biggr].
 \label{Fzosc-integral}
\end{multline}

\begin{figure}
 \centerline{\includegraphics[width=0.95\columnwidth]{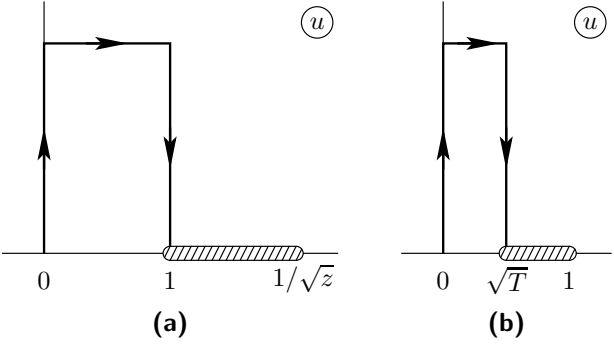}}
 \caption{Integration contours used to calculate oscillating correction to
the (a) generating function [Eq.\ (\protect\ref{dF})] and (b) transmission
distribution [Eq.\ (\protect\ref{PTosc-integral})].}
 \label{Fig:contour}
\end{figure}

The first term of the above expression gives the main contribution to the
generating function, Eq.\ (\ref{Fz}). The next term is suppressed due to
oscillations of the integrand. To estimate this contribution to ${\mathcal
F}(z)$ we will apply a saddle-point method. Representing $\cos(2 u \epsilon L)$
as a real part of an exponential function and deforming the integration contour
as shown in Fig.\ \ref{Fig:contour}a, we get
\begin{equation}
 \delta \mathcal{F}(z)
  = \frac{2 W \epsilon}{\pi} \mathop{\mathrm{Re}}
    \int \frac{u^2 \bigl(u \sqrt{1 - z} - \sqrt{1-z u^2} \bigr)^2\, du}
    {\sqrt{(1 - z)(1 - u^2)^3 (1- z u^2)}}
    e^{2 i u \epsilon L}.
 \label{dF}
\end{equation}
The integrand decays exponentially when the contour runs far from the real axis.
Hence we can estimate the value of the integral by expanding the pre-exponential
factor near the two ends of the contour. We parameterize these two parts by
substituting $u = i w$ and $u = 1 + i w$ and obtain the two contributions
\begin{equation}
 \delta \mathcal{F}_a
  = -\frac{4 W \epsilon}{\pi} \int_0^\infty dw\, w^3 e^{-2 w \epsilon L}
  = -\frac{3 W \epsilon}{2 \pi (\epsilon L)^4},
\end{equation}
\begin{multline}
 \delta \mathcal{F}_b
  = \frac{W \epsilon \sin(2 \epsilon L - \pi/4)}{\sqrt{2} \pi (1 - z)^2}
    \int_0^\infty dw\, \sqrt{w}\, e^{- 2 w \epsilon L} \\
  = \frac{W \epsilon \sin(2 \epsilon L - \pi/4)}
    {8 \sqrt{\pi} (1 - z)^2 (\epsilon L)^{3/2}}.
 \label{Fz2}
\end{multline}
We see that the vicinity of $u = 0$ yields much smaller correction,
$\delta \mathcal{F}_a\ll \delta \mathcal{F}_b$, and hence should
be discarded even though $\delta \mathcal{F}_b$ oscillates.
The generating function including the first oscillating
correction is the sum of Eqs.\ (\ref{Fz}) and (\ref{Fz2}),
\begin{equation}
 \mathcal{F}(z)
   = W \epsilon \left[
       \frac{K(z) - E(z)}{\pi z \sqrt{1-z}}
       +\frac{\sin(2 \epsilon L - \pi/4)}
       {8 \sqrt{\pi} (1 - z)^2 (\epsilon L)^{3/2}}
     \right].
 \label{Fzosc}
\end{equation}
The conductance and the Fano factor [Eqs.\ (\ref{Gosc}) and (\ref{Fosc})] are then
calculated by expanding $\mathcal{F}(z)$ in small $z$.

Let us now calculate the oscillating correction to the distribution function
(\ref{PTelarge}). The second term in Eq.\ (\ref{Fzosc}) for the generating
function does not possess a branch cut at $z>1$ [this is an artifact of the
saddle-point approximation applied to Eq.\ (\ref{dF})]. Therefore, we cannot
get the result by applying Eq.\ (\ref{P-F}) to the generating function
(\ref{Fzosc}). Instead, we have to use a more general expression
(\ref{Fzosc-integral}), which still possesses a branch cut. Performing the
analytic continuation of the integrand in Eq.\ (\ref{Fzosc-integral}) and
using Eq.\ (\ref{P-F}), we obtain the correction to the distribution function in
the form
\begin{equation}
 \delta P(T)
  = \frac{2 W \epsilon}{\pi^2 T^2} \int_0^{\sqrt{T}}
    \frac{du\, u^2 [T - (2 - T) u^2] \cos(2 u \epsilon L)}
    {\sqrt{(1 - T) (1 - u^2)^3 (T - u^2)}}.
 \label{PTosc-integral}
\end{equation}
This integral contains rapidly oscillating function and we calculate it using
the same method as above. We replace $\cos(2 u \epsilon L)$ by exponential,
then deform the contour as shown in Fig.\ \ref{Fig:contour}b,
and estimate the integral in the vicinity of
$u = \sqrt{T}$ by changing variable $u = \sqrt{T} + i w$. The result of this
calculation is
\begin{multline}
 \delta P(T)
  = -\frac{\sqrt{2} W \epsilon \sin(2 \epsilon L \sqrt{T} + \pi/4)}
    {\pi^2 T^{1/4} (1 - T)} \int_0^\infty \frac{dw}{\sqrt{w}}\,
    e^{-2 \epsilon L w} \\
  = -\frac{W \epsilon \sin(2 \epsilon L \sqrt{T} + \pi/4)}
    {\pi^{3/2} T^{1/4} (1 - T) \sqrt{\epsilon L}}.
 \label{PTosccorr}
\end{multline}
Combining Eq.\ (\ref{PTosccorr}) with the main part Eq.\ (\ref{PTelarge}), we
have the following distribution function
\begin{equation}
 P(T)
  = W \epsilon \left[
      \frac{K(T) - E(T)}{\pi^2 T \sqrt{1 - T}}
      -\frac{\sin(2 \epsilon L \sqrt{T} + \pi/4)}
      {\pi^{3/2} T^{1/4} (1 - T) \sqrt{\epsilon L}}
    \right].
 \label{PTosc}
\end{equation}

The correction to the distribution function, Eq.\ (\ref{PTosccorr}), is not
integrable at the point $T = 1$. This prevents us from calculating corrections
to conductance and higher moments using Eq.\ (\ref{PTosc}). In fact, the result
(\ref{PTosc}) is not accurate when $T$ is close to $1$. Indeed, expanding the
integrand in Eq.\ (\ref{PTosc-integral}) near $u = \sqrt{T}$, we have neglected
the variation of the factor $(1 - u^2)^{-3/2}$. When $T$ approaches $1$, this
neglection is not justified because the singularity at $u = 1$ gets close to the
integration contour [see Fig.\ \ref{Fig:contour}b]. The integral in Eq.\
(\ref{PTosccorr}) converges at $w \sim 1/\epsilon L$. When typical values of $w$
are of the order of $1 - \sqrt{T}$ our approach fails. Thus we have to impose
the condition
\begin{equation}
 1 - T
  \gg \frac{1}{\epsilon L}
\end{equation}
for applicability of the result (\ref{PTosc}). This condition also ensures that
the oscillating correction is smaller than the main term in Eq.\ (\ref{PTosc}).

\section{Derivation of the two-loop beta function for random potential}
\label{App:twoloop}

\subsection{The model and universality}

In this Appendix we derive the RG equations for random potential disorder. The
beta function is universal when it is invariant under small changes in the
definition of the coupling constants. Generally, the latter depends on details
of the high-energy part of the spectrum (where the dispersion is no longer
linear) and hence on the way the ultraviolet cutoff of the effective low-energy
theory is imposed. Therefore, the invariance of the beta function with respect
to uncertainty in the definition of couplings is equivalent to its independence
on the RG regularization scheme.

When only one coupling constant is present in the model, the beta function is
universal within the two-loop accuracy. Indeed, in this case a small change in a
coupling constant would only change the three-loop and higher terms in the
beta function. In order to see this, one can assume that the beta function for
some coupling $\alpha$ is known within the three-loop accuracy,
\begin{equation}
 \frac{\partial\alpha}{\partial \ln\Lambda}
  = A \alpha^2 + B \alpha^3 + C \alpha^4,
 \label{alphadot}
\end{equation}
with the coefficients $A$, $B$, and $C$ in the first, second, and third-loop
terms, respectively. Introducing a new coupling $\alpha'$ through
\begin{equation}
 \alpha'
  = \alpha + M \alpha^2 + N \alpha^3,
 \label{alphaprime}
\end{equation}
and using Eq.\ (\ref{alphadot}), one finds the RG equation for this new coupling
in the form
\begin{equation}
 \frac{\partial\alpha'}{\partial \ln\Lambda}
  = A {\alpha'}^2 + B {\alpha'}^3 + (C - A M^2 + A N - B M) {\alpha'}^4.
 \label{alphaprimedot}
\end{equation}
One sees that the coefficients of the first and second terms of the new
beta function remain unchanged, whereas the coefficient in the last term (third
loop) depends on the definition of $\alpha'$.

When the model contains more than one coupling constant, already the two-loop
RG equations are in general non-universal. This can be seen from
\begin{align}
 \frac{\partial\alpha_i}{\partial \ln\Lambda}
  &= A_i^{kl} \alpha_k \alpha_l + B_i^{klm} \alpha_k \alpha_l \alpha_m,
 \label{nonuniv1} \\
 \alpha'_i
  &= \alpha_i + M_i^{kl} \alpha_k \alpha_l, \label{nonuniv2} \\
 \frac{\partial\alpha'_i}{\partial \ln\Lambda}
  &= A_i^{kl} \alpha'_k \alpha'_l \notag \\
     &+\left(B_i^{klm} +2A_j^{kl} M_i^{jm}- 2A_i^{kj} M_j^{lm} \right)
     \alpha'_k \alpha'_l  \alpha'_m. \label{nonuniv3}
\end{align}
In our model, the RG equation for $\alpha_0$ is independent of energy
$\epsilon$, therefore we can retain the two-loop term. On the other hand, the RG
equation for energy involves both $\epsilon$ and $\alpha_0$; hence, we only keep
the one-loop term in the corresponding scaling function.

We choose the dimensional regularization (with the minimal subtraction) as our
RG scheme: \cite{Peskin} we consider the action in $d = 2 - \varepsilon$
dimensions ($\varepsilon > 0$) and send $\varepsilon\to 0$ at the end. The model
is characterized by two constants, a mass parameter $m$ which corresponds to
the imaginary (Matsubara) energy and the coupling constant $\alpha_0$ which
corresponds to the mean quadratic potential disorder strength Eq.\
(\ref{alpha-mu}).

The renormalized action of the model (known as the massive Gross-Neveu model)
has the form 
\begin{equation}
 S_R[\psi]
  = \int d^{2-\varepsilon}x \left[
      \bar\psi \bm{\sigma}\nabla \psi
      +m \bar\psi \psi
      +\pi\alpha_0\mu^\varepsilon (\bar\psi \psi)^2
    \right].
 \label{ActionRG}
\end{equation}
Here $m =-i \epsilon$ and we have introduced the mass scale $\mu$ to fix the
dimension of coupling $\alpha_0$ to zero. The wavefunction $\psi$ is a
$d$-dimensional spinor in the left/right-moving space, $\bm{\sigma}$ is a vector
of $\sigma_\alpha$, which are the generators of the $d$-dimensional Clifford
algebra, obeying
\begin{equation}
 \sigma_\alpha \sigma_\beta + \sigma_\beta \sigma_\alpha
  = 2 \delta_{\alpha\beta}\ \mathbb 1,
\qquad \sum_{\alpha} \sigma_\alpha \sigma_\alpha
= d\; \mathbb 1.
\end{equation}
Our goal is to derive the RG equations for $\alpha_0$ and $m$, which determine
the evolution of these two parameters upon increasing the (infrared) scale of
the model. This derivation closely follows that of Refs.\ \onlinecite{Bondi,
Kivel}, where a related massless theory was considered.

We will start with the one-loop calculation, since the corresponding integrals
and counterterms will be required for the two-loop calculation as well. As
discussed in Sec.\ \ref{Sec:RG}, we will discard diagrams with closed fermionic
loops, as is appropriate for a system with quenched disorder. (Alternatively,
the same result is obtained by using the replica trick or the supersymmetry.)

The quadratic (clean) part of the action yields the bare fermion propagator
(solid line in diagrams),
\begin{equation}
 G^{(0)}(\mathbf{p})
  = (i m - \bm{\sigma} \mathbf{p})^{-1}
  = \frac{-\bm{\sigma} \mathbf{p} - i m}{p^2 + m^2}.
 \label{G}
\end{equation}
Dashed lines in the diagrams denote the disorder correlator
\begin{equation}
 \Gamma{(0)}
  = 2\pi \mu^\varepsilon \alpha_0.
\end{equation}
In order to keep track of the
two-sided algebra structure, we draw the diagrams for vertex corrections with an
upper and lower electron line.

The counterterms are denoted with crossed circles. The vertex counterterm is
\begin{equation}
 \delta \Gamma
  = 2\pi \mu^\varepsilon \delta\alpha_0\; \mathbb{1}\otimes\mathbb{1}
\end{equation}
and the self-energy (mass and velocity) counterterm is
\begin{equation}
 \delta\Sigma
  = (-i \delta m + \delta v_0 \, \bm{\sigma} \mathbf{p})\;
    \mathbb{1}\otimes\mathbb{1}.
\end{equation}
Below we will calculate one- and two-loop diagrams for the self energy and the
vertex amplitude and construct the corresponding counterterms using the minimal
subtraction scheme.

The divergent parts of all the integrals appear with the factor $c_\varepsilon$
or $c_\varepsilon^2$, where
\begin{equation}
 c_\varepsilon
  = (4\pi)^{\tfrac{\varepsilon}{2} - 1} \left(\frac{\mu}{m}\right)^\varepsilon
    \left(1 - \frac{\gamma \varepsilon}{2}\right)
\end{equation}
and $\gamma \approx 0.577$ is the Euler-Mascheroni constant.

\subsection{One-loop RG equations}

We calculate the one-loop integrals with the accuracy $\mathcal O(\varepsilon)$,
because this precision is needed for the two-loop calculation later. The
following integrals appear in the one-loop diagrams of Fig.\ \ref{Fig:oneloop}
(we include $\mu^\varepsilon$ to make the integrals dimensionless):
\begin{align}
 \mu^\varepsilon \int \frac{d^{2-\varepsilon} p}{(2\pi)^{2-\varepsilon}}\,
 \frac{1}{p^2 + m^2}
  &= c_\varepsilon \left[
      \frac{2}{\varepsilon} + \mathcal O(\varepsilon)
    \right], \label{1la}\\
 \mu^\varepsilon \int \frac{d^{2-\varepsilon} p}{(2\pi)^{2-\varepsilon}}\,
 \frac{p_\alpha p_\beta }{(p^2 + m^2)^2}
  &= \delta_{\alpha\beta}\, c_\varepsilon \left[
      \frac{1}{\varepsilon}+ \mathcal O(\varepsilon)
    \right], \label{1lb}\\
 \mu^\varepsilon \int \frac{d^{2-\varepsilon} p}{(2\pi)^{2-\varepsilon}}\,
 \frac{m^2}{(p^2 + m^2)^2}
  &= c_\varepsilon \big[ 1+ \mathcal O(\varepsilon) \big]. \label{1lc}
\end{align}
All other integrals appearing in the one-loop diagrams  are zero
because of isotropy.

Only two diagrams (Fig.\ \ref{Fig:oneloop}a and \ref{Fig:oneloop}b) give the
first-loop corrections to the self energy and vertex, the third and fourth
diagrams (Fig.\ \ref{Fig:oneloop}c and \ref{Fig:oneloop}d) cancel each other
[up to $\mathcal O (1)$]. More specifically, the diagram (c) on its own,
\begin{equation}
 \text{(c)}
  = 4\pi^2 \alpha_0^2 c_\varepsilon \left[
      \mathbb 1\otimes \mathbb 1
      -\frac{1}{\varepsilon} \sum_\alpha \sigma_\alpha\otimes \sigma_\alpha
    \right] + \mathcal O(\varepsilon).
\end{equation}
would generate a new algebraic structure (corresponding to a new disorder ---
random vector potential). We have to calculate it together with its crossed
companion, the diagram (d):
\begin{equation}
 \text{(d)}
  = 4\pi^2 \alpha_0^2 c_\varepsilon \left[
      \mathbb 1\otimes \mathbb 1
      +\frac{1}{\varepsilon} \sum_\alpha \sigma_\alpha\otimes \sigma_\alpha
    \right] + \mathcal O(\varepsilon).
\end{equation}
In combination of the two diagrams, the new structure is canceled. This happens
also in higher loops, where all new algebraic structures always cancel. We
therefore combine diagrams with their crossed versions directly.

The one-loop corrections read
\begin{align}
 \Sigma^{(1)}\big|_{p=0}
  &= \text{(a)}
  = -2\pi i m\, \alpha_0\,
    \mu^\varepsilon \int \frac{d^{2-\varepsilon} p}{(2\pi)^{2-\varepsilon}p}
    \frac{1}{p^2 + m^2} \notag\\
  &= -2\pi i m\, \alpha_0 c_\varepsilon \left[
      \frac{2}{\varepsilon} + \mathcal O (\varepsilon)
    \right], \label{eqsigma1l}\\
 \Gamma^{(1)}\big|_{p=0}
  &= 2 \times \text{(b)} + \text{(c)} + \text{(d)} \notag\\
  &= 8\pi^2 \alpha_0^2\, \sigma_\alpha \sigma_\beta\,
    \mu^{2\varepsilon} \int \frac{d^{2-\varepsilon}p}{(2\pi)^{2-\varepsilon}}
    \frac{p_\alpha p_\beta}{(p^2 + m^2)^2} \notag\\
  &\quad - 8\pi^2 \alpha_0^2\,
    \mu^{2\varepsilon} \int \frac{d^{2-\varepsilon}p}{(2\pi)^{2-\varepsilon}}
    \frac{m^2}{(p^2 + m^2)^2} + \mathcal O(1) \notag\\
  &= 16\pi^2 \alpha_0^2 \mu^\varepsilon c_\varepsilon \left[
      \frac{1}{\varepsilon} + \mathcal O(1)
    \right].
 \label{eqgamma1l}
\end{align}
The one-loop correction to the self energy is independent of external momenta,
and therefore, there is no one-loop correction to the velocity. Thus no
rescaling of the fields is required.

Within the minimal subtraction scheme, the divergences in Eqs.\
\eqref{eqsigma1l} and \eqref{eqgamma1l} are canceled by the following one-loop
counterterms:
\begin{equation}
 \delta^{(1)} m
  = -\frac{m\alpha_0}{\varepsilon},
 \qquad\qquad
 \delta^{(1)} \alpha_0
  = -\frac{2\alpha_0^2}{\varepsilon},
 \label{ct1}
\end{equation}
yielding the one-loop RG equations
\begin{equation}
 \frac{\partial \alpha_0}{\partial \ln \Lambda}
  = 2 \alpha_0^2,
 \qquad
 \frac{\partial \epsilon}{\partial \ln \Lambda}
  = \epsilon \alpha_0,
 \label{eq1lrg}
\end{equation}
with $\Lambda$ being the real-space running scale ($\Lambda\sim \mu^{-1}$) and
energy $\epsilon = i m$.

\subsection{Second loop}

Let us now turn to the second-loop calculation. In addition to the integrals
appearing in the one-loop diagrams, the further two single-integrals appear in
the two-loop calculation:
\begin{align}
 \mu^\varepsilon \int \frac{d^{2-\varepsilon}p}{(2\pi)^{2-\varepsilon}}\,
 \frac{m^4}{(p^2 + m^2)^3}
  &= c_\varepsilon \left[
      \frac{1}{2} + \mathcal O(\varepsilon)
    \right], \label{1lo1a}\\
 \mu^\varepsilon \int \frac{d^{2-\varepsilon}p}{(2\pi)^{2-\varepsilon}}\,
 \frac{m^2\, p_\alpha p_\beta}{(p^2 + m^2)^3}
  &= \delta_{\alpha\beta}\, c_\varepsilon \left[
      \frac{1}{4} + \mathcal O(\varepsilon)
    \right]. \label{1lo1b}
\end{align}
The two-loop calculation will only be performed up to $\mathcal O(1)$, since
only the divergent parts of the diagrams are required for deriving the RG
equations.

The double integrals that appear in the calculation of the second-loop diagrams
can be reduced to the following set:
\begin{align}
 &\mu^{2\varepsilon} \int \frac{d^d p\, d^d q}{(2\pi)^{2d}}
 \frac{p_\alpha p_\beta}{(p^2 + m^2) (q^2 + m^2) [(p + q)^2 + m^2]} \notag\\
  &\qquad\qquad = \delta_{\alpha \beta} \, c_\varepsilon^2 \left[ 
      \frac{2}{\varepsilon^2} + \frac{1}{\varepsilon} + \mathcal O (1)
    \right], \label{2l3a}\\
 &\mu^{2\varepsilon} \int \frac{d^d p\, d^d q}{(2\pi)^{2d}}
 \frac{p_\alpha q_\beta}{(p^2 + m^2) (q^2 + m^2) [(p + q)^2 + m^2]} \notag\\
  &\qquad\qquad = \delta_{\alpha \beta}\, c_\varepsilon^2 \left[
      -\frac{1}{\varepsilon^2} - \frac{1}{2\varepsilon} + \mathcal O (1)
    \right], \label{2l3b} \\
 &\mu^{2\varepsilon} \int \frac{d^d p\, d^d q}{(2\pi)^{2d}}
 \frac{p_\alpha q_\beta (p + q)_\mu (p + q)_\nu}
 {(p^2 + m^2) (q^2 + m^2) [(p + q)^2 + m^2]^2} \notag\\
  &= c_\varepsilon^2 \left[
      -\frac{\delta_{\alpha \beta} \delta_{\mu \nu}}{2 \varepsilon^2}
      +\frac{\delta_{\alpha \mu}\delta_{\beta \nu}
             +\delta_{\alpha \nu}\delta_{\beta \mu}
             -\delta_{\alpha \beta}\delta_{\mu \nu} }{8 \varepsilon}
    \right] + \mathcal O(1), \label{2l4a} \\
 &\mu^{2\varepsilon} \int \frac{d^d p\, d^d q}{(2\pi)^{2d}}
 \frac{m^2 p_\alpha p_\beta}
 {(p^2 + m^2) (q^2 + m^2) [(p + q)^2 + m^2]^2} \notag\\
  &\qquad\qquad = \delta_{\alpha \beta}\, c_\varepsilon^2 \left[
      \frac{1}{\varepsilon} + \mathcal O (1)
    \right], \label{2l4b}\\
 &\mu^{2\varepsilon} \int \frac{d^d p\, d^d q}{(2\pi)^{2d}}
 \frac{m^2 p_\alpha q_\beta}
 {(p^2 + m^2) (q^2 + m^2) [(p + q)^2 + m^2]^2} \notag\\
  &\qquad\qquad = \delta_{\alpha \beta}\, c_\varepsilon^2 \left[
      -\frac{1}{\varepsilon} + \mathcal O (1)
    \right]. \label{2l4c}
\end{align}
The integrals with $m^2$ and $m^4$ in the numerator of an integrand of the type
Eqs. \eqref{2l3a} -- \eqref{2l3b} and \eqref{2l4a} -- \eqref{2l4c},
respectively, are not divergent [$\sim \mathcal O (1)$]. All other integrals
appearing in the diagrams can be derived from the above set by combining
\eqref{2l3a} -- \eqref{2l4c} and/or using $p\leftrightarrow q$.

\subsubsection{Self energy}

\begin{figure*}
 \centerline{\includegraphics[width=\textwidth]{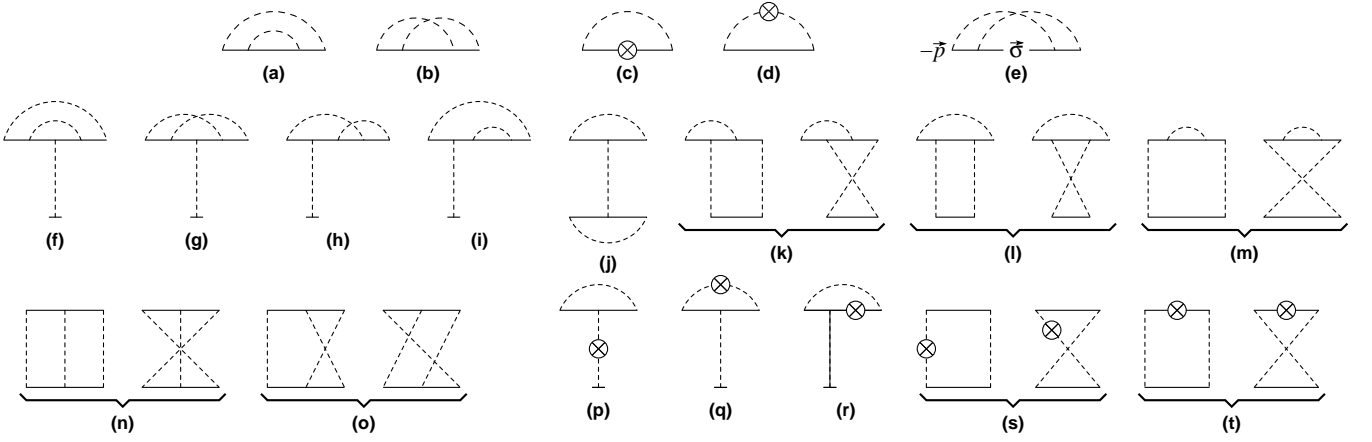}}
 \caption{Two-loop diagrams for the self energy (a--e) and vertex (f--t)
corrections.}
 \label{Fig:2loopdiagrams}
\end{figure*}

\begin{table}
\caption{Two-loop diagrams. The first column contains combinatorial factors and
diagram labels according to Fig.\ \protect\ref{Fig:2loopdiagrams}. The integrals
relevant for the calculation of a given diagram are listed in the second
column. The third column contains the divergent part of the diagram.}
\label{tabdiagrams}
\begin{tabular}{ccc}
\hline\hline
 diagram & integrals used & result \\\hline\\[-10pt]
\multicolumn{3}{c}{
 self energy
 [in units $4\pi^2 i m \alpha_0^2 c_\varepsilon^2/\varepsilon^2$]
}\\[3pt]
(a) & \eqref{1la}$\cdot$\eqref{1lb}, \eqref{1la}$\cdot$\eqref{1lc}
    & $-4 + 4 \varepsilon$\\
(b) & \eqref{2l3a}, \eqref{2l3b}
    & $-2$\\\hline\\[-10pt]
\multicolumn{3}{c}{
 self energy with counterterms
 [in units $4\pi^2 i m \alpha_0^2 c_\varepsilon/4\pi\varepsilon^2$]
}\\[3pt]
(c) & \eqref{1lb}
    & $4 - 4\varepsilon$\\
(d) & \eqref{1la}
    & $8$ \\\hline\\[-10pt]
\multicolumn{3}{c}{velocity} \\[3pt]
(e) &\eqref{2l4a}, \eqref{2l4b}
    & $-4\pi^2 \bm{\sigma}\mathbf{p}\, \alpha_0^2 c_\varepsilon^2/\varepsilon$
      \\\hline\hline\\[-10pt]
\multicolumn{3}{c}{
 vertex
 [in units $8\pi^3 \alpha_0^3 \mu^\varepsilon c_\varepsilon^2/\varepsilon^2$]
}\\[3pt]
2$\times$(f) &\eqref{1lb}$^2$, \eqref{1lb}$\cdot$\eqref{1lc}
    & $8 - 8\varepsilon - \langle 8\varepsilon \rangle$\\
2$\times$(g) &\eqref{2l4a}, \eqref{2l4c}
    & $-4 + 4\varepsilon + \langle 4\varepsilon \rangle$\\
4$\times$(h) &\eqref{2l4a}, \eqref{2l4c}
    & $8 - 8\varepsilon -\langle 8\varepsilon \rangle$\\
4$\times$(i)
    & \eqref{1la}$\cdot$\eqref{1lo1a}, \eqref{1la}$\cdot$\eqref{1lo1b}
    & $-\langle 8\varepsilon \rangle$\\
(j) & \eqref{1lb}$^2$, \eqref{1lb}$\cdot$\eqref{1lc}
    & $4 - 4\varepsilon - \langle 4\varepsilon \rangle$\\
4$\times$(k) & \eqref{2l4c}
    & $-\langle 16\varepsilon \rangle$\\
2$\times$(l) & \eqref{2l4c}
    & $0$\\
2$\times$(m)
    & \eqref{1la}$\cdot$\eqref{1lo1a}, \eqref{1la}$\cdot$\eqref{1lo1b}
    & $0$\\
(n) & \eqref{1lb}$^2$
    & $4 - 2\varepsilon$\\
2$\times$(o) & \eqref{2l4a}
    & $-4 + 4\varepsilon$\\\hline\\[-10pt]
\multicolumn{3}{c}{
 vertex with counterterms
 [in units $8\pi^3 \alpha_0^3 \mu^\varepsilon c_\varepsilon/4\pi\varepsilon^2$]
}\\[3pt]
2$\times$(p) & \eqref{1lb}
    & $-16 + 8\varepsilon + \langle 8\varepsilon \rangle$\\
2$\times$(q) & \eqref{1lb}
    & $-16 + 8\varepsilon + \langle 8\varepsilon \rangle$\\
4$\times$(r) & \eqref{1lo1a}, \eqref{1lo1b}
    & $\langle 8\varepsilon \rangle$\\
2$\times$(s) & \eqref{1lc}
    & $\langle 16\varepsilon \rangle$\\
2$\times$(t) & \eqref{1lo1a}, \eqref{1lo1b}
    & $0$
\\\hline\hline
\end{tabular}
\end{table}

Let us calculate corrections to the self energy in two-loop order (including
diagrams with the one-loop counterterms). The relevant diagrams are shown in
Fig.\ \ref{Fig:2loopdiagrams}a -- \ref{Fig:2loopdiagrams}e. The divergent
parts of these diagrams and the integrals used in their calculation are given in
Table \ref{tabdiagrams}. The divergent part of the self-energy correction is
\begin{equation}
 \Sigma^{(2)}\big|_{p=0}
  = \frac{3 i}{2}\, \frac{m \alpha_0^2}{\varepsilon^2}.
\end{equation}
The divergence in $\Sigma^{(2)}$ is compensated by the second-order mass
counterterm
\begin{equation}
 \delta^{(2)} m
  = \frac{3}{2}\, \frac{m \alpha_0^2}{\varepsilon^2}.
\end{equation}

Electron velocity $v_0$ also acquires a correction in two-loop order. This
correction appears because the second-order self energy depends on the external
momentum. Expanding $\Sigma^{(2)}$ in small external momentum we obtain the
diagram Fig.\ \ref{Fig:2loopdiagrams}e. It equals to
\begin{equation}
 \text{(e)}
  = \mathbf{p}\, \frac{\partial \Sigma^{(2)}(\mathbf{p})}{\partial \mathbf{p}}
    \bigg|_{p=0}
  = \bm{\sigma} \mathbf{p}\, 4\pi^2 \alpha_0^2 c_\varepsilon^2 \biggl[
      -\frac{1}{\varepsilon} + \mathcal O (1)
    \biggr].
\end{equation}
This divergence is compensated by the velocity counterterm
\begin{equation}
 \delta^{(2)} v_0
  = \frac{\alpha_0^2}{4\varepsilon}.
\end{equation}

\subsubsection{Vertex}

The two-loop vertex diagrams are shown in Fig.\ \ref{Fig:2loopdiagrams}f --
\ref{Fig:2loopdiagrams}t. The values of these diagrams are given in the bottom
part of Table \ref{tabdiagrams} along with the integrals used in their
calculation. According to Refs.\ \onlinecite{Bondi, Kivel}, one can disregard
the mass in the numerator of the electron propagator \eqref{G}. In order to
check this fact, we keep the corresponding parts in angular brackets in Table
\ref{tabdiagrams}. Indeed, these contributions sum up to zero.

The divergent part of the two-loop vertex correction is
\begin{multline}
 \Gamma^{(2)}\big|_{p=0}
  = 8\pi^3 \alpha_0^3 \mu^\varepsilon \left(
      \frac{16 c_\varepsilon^2}{\varepsilon^2}
      -\frac{32 c_\varepsilon}{4\pi \varepsilon^2}
      +\frac{2c_\varepsilon^2}{\varepsilon}
    \right) \\
  = \pi \mu^\varepsilon \alpha_0^3 \left(
      -\frac{8}{\varepsilon^2}
      +\frac{1}{\varepsilon}
    \right).
\end{multline}
This divergence is canceled by the two-loop vertex counterterm
\begin{equation}
 \delta^{(2)} \alpha_0
  = \alpha_0^3 \left(
      \frac{4}{\varepsilon^2} - \frac{1}{2\varepsilon}
    \right).
\end{equation}

\subsection{RG equations}

Now we collect the one and two-loop counterterms and compose the bare action of
the model:
\begin{multline}
 S_B
  = \!\int\! d^{2-\varepsilon} x \big[
      \bar\psi_B \bm{\sigma}\nabla \psi_B
      +m_B \bar\psi_B \psi_B 
      +\pi\alpha_{0B} (\bar\psi_B \psi_B)^2
    \big] \\
  = \!\int\! d^{2-\varepsilon}x \bigg[
      \left(
        1 + \frac{\alpha_0^2}{4\varepsilon}
      \right) \bar\psi \bm{\sigma}\nabla \psi 
      +m \left(
        1 - \frac{\alpha_0}{\varepsilon} + \frac{3\alpha_0^2}{2\varepsilon^2}
      \right) \bar\psi \psi \\
      +\pi \alpha_0 \mu^\varepsilon \left(
        1 - \frac{2\alpha_0}{\varepsilon} + \frac{4\alpha_0^2}{\varepsilon^2}
        -\frac{\alpha_0^2}{2\varepsilon}
      \right) (\bar\psi \psi)^2
    \bigg].
\end{multline}
Parameters of this action are
\begin{align}
 \psi_B
  &= \psi \left(
      1 + \frac{\alpha_0^2}{8\varepsilon}
    \right), \\
 m_B
  &= m \left(
      1 - \frac{\alpha_0}{\varepsilon} + \frac{3\alpha_0^2}{2\varepsilon^2}
      -\frac{\alpha_0^2}{4\varepsilon}
    \right), \\
  \alpha_{0B}
  &= \alpha_0 \mu^\varepsilon \left(
      1 - \frac{2\alpha_0}{\varepsilon} + \frac{4\alpha_0^2}{\varepsilon^2}
      -\frac{\alpha_0^2}{\varepsilon}
    \right).
\end{align}
By construction, the bare couplings $m_B$ and $\alpha_{0B}$ do not depend on the
scale $\mu$,
\begin{equation}
 \frac{\partial \alpha_{0B}}{\partial \ln \mu}
  = 0,
 \qquad
 \frac{\partial m_B}{\partial \ln \mu}
  = 0.
\end{equation}
This determines scaling behavior of renormalized (observable) couplings:
\begin{align}
 \frac{\partial \alpha_0}{\partial \ln \mu}
  &= -\varepsilon \alpha_0 - 2\alpha_0^2 - 2 \alpha_0^3, \\
 \frac{\partial m}{\partial \ln \mu}
  &= -m \alpha_0 - \frac{m\alpha_0^2}{2}.
\end{align}
We express the result in the form of real-space scaling with $\Lambda \sim
\mu^{-1}$. This amounts to changing the sign of the derivatives. Taking the
limit $\varepsilon \to 0$ and replacing mass by the energy ($m = -i \epsilon$),
we finally obtain the RG equations in the form
\begin{align}
 \frac{\partial \alpha_0}{\partial \ln \Lambda}
  &= 2\alpha_0^2 + 2 \alpha_0^3, \label{alphatwoloop}
 \\
 \frac{\partial \epsilon}{\partial \ln \Lambda}
  &= \epsilon \left( \alpha_0 + \frac{\alpha_0^2}{2} \right).
 \label{Etwoloop}
\end{align}

As discussed above [see Eqs.\ (\ref{nonuniv1}) -- (\ref{nonuniv3})], the
second-loop term in the RG equation for energy \eqref{Etwoloop} is not
universal. This is easily demonstrated with the help of
a small redefinition of $\alpha_0$,
\begin{align}
 \alpha_0'
  &= \alpha_0 + M \alpha_0^2, \\
 \frac{\partial \epsilon}{\partial \ln \Lambda}
  &= \epsilon \left(
      \alpha_0' + \frac{{\alpha_0'}^2}{2} -M{\alpha_0'}^2
    \right).
\end{align}
On the contrary, the two-loop term in the beta function for $\alpha_0$ remains
unchanged as the RG equation \eqref{alphatwoloop}, being dimensionless, cannot
contain $\epsilon$. In the main text, we use only universal one-loop part of
the energy beta function.

\section{Analysis of one-loop RG equations}
\label{App:oneloop}

In this Appendix we will analyze the set of coupled RG equations (\ref{RGa0}),
(\ref{RGaperp}) and (\ref{RGaz})
assuming that all three couplings are nonzero.
It is convenient to introduce new couplings
\begin{eqnarray}
\alpha_1&=&\alpha_\perp+\alpha_z~, \label{alpha1}\\
\alpha_2&=&\alpha_\perp-2\alpha_z~. \label{alpha2}
\end{eqnarray}
In terms of these couplings the one-loop RG equations take the form
\begin{align}
 \frac{\partial\alpha_0}{\partial \ln \Lambda}
  &= 2 \alpha_0^2 +2\alpha_0\alpha_1+\frac{2}{9}(2\alpha_1 + \alpha_2)(\alpha_1 - \alpha_2), \label{RGnewa0}\\
 \frac{\partial\alpha_1}{\partial \ln \Lambda}
  &
  = 2 \alpha_0 \alpha_1 + \frac{2}{9}(\alpha_1 + 2\alpha_2)(\alpha_1 - \alpha_2), \label{RGnewa1}\\
 \frac{\partial\alpha_2}{\partial \ln \Lambda}
  &= -4\alpha_0\alpha_2 - \frac{4}{9}(\alpha_1 + 2\alpha_2)(\alpha_1 - \alpha_2).  \label{RGnewa2}
\end{align}
It follows that the couplings $\alpha_0$ and $\alpha_1$ grow upon the renormalization.
Furthermore, comparing Eqs.\ (\ref{RGnewa0}) and (\ref{RGnewa1}), one sees that
the coupling $\alpha_0$ increases faster than $\alpha_1$. On the other hand,
the coupling $\alpha_2$ decreases, which allows us to neglect $\alpha_2$ in
Eqs.\ (\ref{RGnewa0}) and (\ref{RGnewa1}) as the first approximation:
\begin{align}
 \frac{\partial\alpha_0}{\partial \ln \Lambda}
  &= 2 \alpha_0^2 +2\alpha_0\alpha_1+\frac{4}{9}\alpha_1^2, \label{RGappra0}\\
 \frac{\partial\alpha_1}{\partial \ln \Lambda}
  &
  = 2 \alpha_0 \alpha_1 + \frac{2}{9}\alpha_1^2. \label{RGappra1}
\end{align}
Let us consider the ratio of the couplings, which we denote
\begin{equation}
g_1=\frac{\alpha_1}{\alpha_0}. \label{g1intro}
\end{equation}
Using Eqs.\ (\ref{RGappra0}) and (\ref{RGappra1}), we obtain the RG equation for
$g_1$ in the form
\begin{equation}
 \frac{\partial g_1}{\partial \ln \Lambda}
  = -\frac{16}{9}\; g_1^2 \left(\alpha_0 + \frac{1}{4} \alpha_1 \right).
 \label{RGg1}
\end{equation}

Assume bare values of the three couplings are of the same order,
$\alpha_0 \sim \alpha_\perp \sim \alpha_z \sim \alpha$.
Since $\alpha_0$ grows faster than $\alpha_1$,  for the ``first iteration" we retain only $\alpha_0$ 
in Eq.\ (\ref{RGg1}). We find
\begin{equation}
 \frac{1}{g_1(\Lambda)}
  \simeq \frac{1}{g_1} + \frac{16}{9} \int_a^{\Lambda}
    \frac{d\tilde\Lambda}{\tilde\Lambda}\; \alpha_0(\tilde\Lambda).
 \label{1g1}
\end{equation}
Further, neglecting $\alpha_1$ also in  Eq.\ (\ref{RGappra0}), dividing this
equation
by $\alpha_0$, and integrating the result over $d\ln\Lambda$, we find
\begin{equation}
 2 \int_a^{\Lambda} \frac{d\tilde\Lambda}{\tilde\Lambda}\;
\alpha_0(\tilde\Lambda)
 = \ln \frac{\alpha_0(\Lambda)}{\alpha}.
\label{ln-ln}
\end{equation}
When $\alpha_0$ reaches unity (this happens when $\epsilon < \gamma$), combining
Eqs.\ (\ref{1g1}) and (\ref{ln-ln}), we have
\begin{equation}
 \frac{1}{g_1(\Lambda)}
  = \frac{1}{\alpha_1}
  \simeq \frac{1}{g_1} + \frac{8}{9}\ln\frac{1}{\alpha}.
\end{equation}
We arrive at the conclusion that at the end of the renormalization the scalar potential
becomes the dominating disorder:
\begin{equation}
 \left. \frac{\alpha_0}{\alpha_\perp + \alpha_z} \right|_{\alpha_0 \simeq 1}
  \simeq \frac{8}{9} \ln\frac{1}{\alpha}
  \gg 1.
 \label{logalpha0}
\end{equation}
This result can be refined by retaining $\alpha_1$ in Eqs.\ (\ref{RGappra0}),
(\ref{RGappra1}), and (\ref{RGg1}), which yields a correction $\simeq -(3/4)
\ln|\ln\alpha|$ to Eq.\ (\ref{logalpha0}).

Above we assumed that all three couplings are of the same order at the ultraviolet scale.
The generalization onto the case of non-equal couplings is straightforward.
It turns out that in the ultraballistic regime (when the renormalization stops by the
largest coupling reaching unity) the scalar potential always wins, whatever the initial couplings are.


\begin{thebibliography}{99}

\bibitem{Novoselov04}
K.S.~Novoselov, A.K.~Geim, S.V.~Morozov, D.~Jiang, Y.~Zhang, S.V.~Dubonos,
I.V.~Grigorieva, and A.A.~Firsov, Science \textbf{306}, 666 (2004).

\bibitem{NovoselovPNAS}
K.S.~Novoselov, D.~Jiang, F.~Schedin, T.J.~Booth, V.V.~Khotkevich, S.V.~Morozov,
and A.K.~Geim, Proc.\ Natl.\ Acad.\ Sci.\ U.S.A.\ \textbf{102}, 10451 (2005).

\bibitem{Novoselov05}
K.S.~Novoselov, A.K.~Geim, S.V.~Morozov, D.~Jiang, M.I.~Katsnelson,
I.V.~Grigorieva, S.V.~Dubonos, and A.A.~Firsov, Nature (London) \textbf{438},
197 (2005).

\bibitem{Zhang05}
Y.~Zhang, Y.-W.~Tan, H.L.~Stormer, and P.~Kim, Nature (London) \textbf{438}, 201
(2005).

\bibitem{Kim}
Y.-W. Tan, Y.~Zhang, H.L.~Stormer, and P.~Kim, Eur.\ Phys.\ J.\ Special Topics,
\textbf{148}, 15 (2007).

\bibitem{Geim07}
A.K.~Geim and K.S.~Novoselov, Nature Materials \textbf{6}, 183 (2007).

\bibitem{RMP07}
A.H.~Castro Neto, F.~Guinea, N.M.R.~Peres, K.S.~Novoselov, and A.K.~Geim,
arXiv:0709.1163, to appear in Rev.\ Mod.\ Phys.

\bibitem{Aleiner06}
I.L.~Aleiner and K.B.~Efetov, Phys.\ Rev.\ Lett.\ \textbf{97}, 236801 (2006).

\bibitem{Altland06}
A.~Altland, Phys.\ Rev.\ Lett.\  \textbf{97}, 236802 (2006).

\bibitem{OurPRB}
P.M.~Ostrovsky, I.V.~Gornyi, and A.D.~Mirlin, Phys.\ Rev.\ B \textbf{74}, 235443
(2006).

\bibitem{OurPRL}
P.M.~Ostrovsky, I.V.~Gornyi, and A.D.~Mirlin, Phys.\ Rev.\ Lett.\ \textbf{98},
256801 (2007).

\bibitem{OurEPJ}
P.M.~Ostrovsky, I.V.~Gornyi, and A.D.~Mirlin,
Eur.\ Phys.\ J. Special Topics \textbf{148}, 63 (2007).

\bibitem{OurQHE}
P.M.~Ostrovsky, I.V.~Gornyi, and A.D.~Mirlin, Phys.\ Rev.\ B \textbf{77}, 195430
(2008).

\bibitem{McCann}
E.~McCann, K.~Kechedzhi, V.I.~Fal'ko, H.~Suzuura, T.~Ando, and B.L.~Altshuler,
Phys.\ Rev.\ Lett.\ \textbf{97}, 146805 (2006).

\bibitem{GuineaMorpurgo} A.F.~Morpurgo and F.~Guinea,
Phys.\ Rev.\ Lett.\ \textbf{97}, 196804 (2006).

\bibitem{Ryu07}
S.~Ryu, C.~Mudry, H.~Obuse, and A.~Furusaki, Phys.\ Rev.\ Lett.\ \textbf{99},
116601 (2007).

\bibitem{Koshino07}
K.~Nomura, M.~Koshino, and S.~Ryu, Phys.\ Rev.\ Lett.\ \textbf{99}, 146806 (2007).

\bibitem{Hasan08}
D.~Hsieh, D.~Qian, L.~Wray, Y.~Xia, Y.~Hor, R.J.~Cava and M.Z.~Hasan
Nature \textbf{452}, 970 (2008).

\bibitem{Schnyder08} A.P.~Schnyder, S.~Ryu, A.~Furusaki, and A.W.W. Ludwig,
arXiv:0803.2786

\bibitem{Nomura07}
K.~Nomura and A.H.~MacDonald, Phys.\ Rev.\ Lett.\ \textbf{98}, 076602 (2007).

\bibitem{Rycerz07} A.~Rycerz, J.~Tworzydlo, and C.W.J.~Beenakker,
Europhys.Lett. \textbf{79}, 57003 (2007).

\bibitem{Bardarson07}
J.H.~Bardarson, J.~Tworzyd\l o, P.W.~Brouwer, and C.W.J.~Beenakker, Phys.\ Rev.\
Lett.\ \textbf{99}, 106801 (2007).

\bibitem{San-Jose07}
P.~San-Jose, E.~Prada, and D.S.~Golubev, Phys.\ Rev.\ B \textbf{76}, 195445
(2007).

\bibitem{Lewenkopf08}  C.H.~Lewenkopf, E.R.~Mucciolo, and A.H.~Castro Neto,
Phys.\ Rev.\ B \textbf{77}, 081410R (2008).

\bibitem{Katsnelson06}
M.I.~Katsnelson, Eur.\ Phys.\ J.\ B \textbf{51}, 157 (2006).

\bibitem{Tworzydlo06}
J.~Tworzydlo, B.~Trauzettel, M.~Titov, A.~Rycerz, and C.W.J.~Beenakker,
Phys.\ Rev.\ Lett.\ \textbf{96}, 246802 (2006).

\bibitem{Morpurgo06}
H.B.~Heersche, P.~Jarillo-Herrero, J.B.~Oostinga, L.M.K.~Vandersypen,
and A.F.~Morpurgo,
Nature \textbf{446}, 56 (2007).

\bibitem{Miao07} F.~Miao, S.~Wijeratne, Y.~Zhang, U.C.~Coskun, W.~Bao,
and C.N.~Lau, Science \textbf{317}, 1530 (2007).

\bibitem{Danneau07}
R.~Danneau, F.~Wu, M.F.~Craciun, S.~Russo, M.Y.~Tomi, J.~Salmilehto,
A.F.~Morpurgo, and P.J.~Hakonen, Phys.\ Rev.\ Lett.\ \textbf{100},
196802 (2008).

\bibitem{Danneau08}
R.~Danneau, F.~Wu, M.F.~Craciun, S.~Russo, M.Y.~Tomi, J.~Salmilehto,
A.F.~Morpurgo, and P.J.~Hakonen, arXiv:0807.0157.

\bibitem{BeenakkerRMP}
 C.W.J.~Beenakker, Rev.\ Mod.\ Phys.\ \textbf{69}, 731 (1997).

\bibitem{Ludwig07}
S.~Ryu, C.~Mudry, A.~Furusaki, and A.W.W.~Ludwig,
Phys.\ Rev.\ B \textbf{75}, 205344 (2007).

\bibitem{foot-1D}
Some analytical results for ballistic transport in graphene were obtained within
the toy-model of one-dimensional disorder in Refs.\ \onlinecite{San-Jose07,
Titov07}. In the present paper, we address the realistic case of a truly
two-dimensional disorder.

\bibitem{Titov07}
M.~Titov, Europhys.\ Lett.\ \textbf{79}, 17004 (2007).

\bibitem{Titov06} M.~Titov and C.W.J.~Beenakker, Phys. Rev. B \textbf{74},
041401(R) (2006).

\bibitem{CheianovFalko}
V.V.~Cheianov and V.I.~Fal'ko, Phys.\ Rev.\ B \textbf{74}, 041403 (2006).

\bibitem{Savchenko}
F.V.~Tikhonenko, D.W.~Horsell, R.V.~Gorbachev, and A.K.~Savchenko, Phys.\ Rev.\
Lett.\ \textbf{100}, 056802 (2008).

\bibitem{Ando06}
T.~Ando, J.\ Phys.\ Soc.\ Jpn \textbf{75}, 074716 (2006).

\bibitem{Nomura06}
K.~Nomura and A.H.~MacDonald, Phys.\ Rev.\ Lett. \textbf{96}, 256602 (2006).

\bibitem{Khveshchenko}
D.V.~Khveshchenko, Phys.\ Rev.\ B \textbf{75}, 241406(R) (2007).

\bibitem{footnote-unitary}
The linear dependence of the conductivity on the gate voltage (with logarithmic
corrections) could also be explained within the model of very strong short-range
scatterers (unitary limit), such as, e.g., vacancies or substitutional defects,
see Ref.\ \onlinecite{OurPRB}. However, this type of disorder is not observed by
microscopy studies and would be in conflict \cite{OurQHE} with the observed
odd-integer quantum Hall effect.

\bibitem{footnote-EFinLeads}
At finite Fermi energy $E_F$ in the contacts, the right- and left-moving
components of $\psi$ get mixed outside the sample. However, this effect
disappears in the limit $E_F \to \infty$.

\bibitem{Dorokhov83}
O.N.~Dorokhov, Zh.\ Eksp.\ Teor.\ Fiz.\ \textbf{85}, 1040 (1983) [Sov.\ Phys.\
JETP \textbf{58}, 606 (1983)].

\bibitem{LeeLevitovYakovets}
H.~Lee, L.S.~Levitov, and A.Yu.~Yakovets, Phys.\ Rev.\ B \textbf{51} 4079
(1995).

\bibitem{footnote-elliptic}
To avoid confusion, we adopt the following convention for elliptic integrals:
\begin{align*}
 K(m)
  &= \int_0^{\pi/2} \frac{d\theta}{\sqrt{1 - m\sin^2\theta}}, \\
 E(m)
  &= \int_0^{\pi/2} d\theta\; \sqrt{1 - m\sin^2\theta}.
\end{align*}
One often uses an alternative notation, $K(k)$ and $E(k)$, with the modulus $k
= \sqrt{m}$ as an argument.

\bibitem{Schomerus07}
H.~Schomerus, Phys.\ Rev.\ B \textbf{76}, 045433 (2007).

\bibitem{Blanter07}
Ya.M.~Blanter and I.~Martin, Phys.\ Rev.\ B \textbf{76}, 155433 (2007).

\bibitem{Dotsenko}
V.S.~Dotsenko and V.S.~Dotsenko, Adv. Phys. \textbf{32}, 129 (1983).

\bibitem{Ludwig}
A.W.W.~Ludwig, M.P.A.~Fisher, R.~Shankar, and G.~Grinstein, Phys. Rev.~B
\textbf{50}, 7526 (1994);

\bibitem{Tsvelik}
A.M.~Tsvelik, Phys.\ Rev.\ B \textbf{51}, 9449 (1995).

\bibitem{NersesyanTsvelik}
A.~A.~Nersesyan, A.~M.~Tsvelik, and F.~Wenger, Phys.\ Rev.\ Lett.\ \textbf{72},
2628 (1994);
Nucl. Phys.~B \textbf{438}, 561 (1995).

\bibitem{Bocquet}
M.~Bocquet, D.~Serban, and M.R.~Zirnbauer, Nucl.\ Phys.\ B \textbf{578},628
(2000).

\bibitem{AltlandSimonsZirnbauer}
A.~Altland, B.D.~Simons, and M.R.~Zirnbauer, Phys.\ Rep.\ \textbf{359}, 283
(2002).

\bibitem{Guruswamy}
S.~Guruswamy, A.~LeClair, and A.W.W.~Ludwig, Nucl.\ Phys.\ B \textbf{583}, 475
(2000).

\bibitem{footnote-normal-dimension}
In Ref.\ \protect\cite{OurPRB}, the beta function for energy contains an extra
term corresponding to the normal dimension of $\epsilon$. In the present work we
adopt an alternative approach with the running band width.

\bibitem{Nazarov94}
Yu.V.~Nazarov, Phys.\ Rev.\ Lett.\ \textbf{73}, 134 (1994).

\bibitem{Nazarov99}
Yu.~Nazarov, Ann.\ Phys.\ (Leipzig) \textbf{8}, 193 (1999).

\bibitem{Gutman03}
D.B.~Gutman, Y.~Gefen, and A.D.~Mirlin,
in "Quantum Noise in Mesoscopic Physics", edited by Yu.V.~Nazarov (Kluwer,
2003), pp.497-524.

\bibitem{Gutman04}
D.B.~Gutman, A.D.~Mirlin, and Y.~Gefen, Phys.\ Rev.\ B {\bf 71}, 085118 (2005).

\bibitem{Fogler}
M.M.~Fogler, F.~Guinea, M.I.~Katsnelson, arXiv:0807.3165.

\bibitem{Titov-private}
M.~Titov, private communication.

\bibitem{MirlinRMP}
F.~Evers and A.D.~Mirlin, arXiv:0707.4378.

\bibitem{footnote-noantilocalization}
When $\alpha_\perp$ or $\alpha_z$ initially dominate and the energy is
sufficiently high, RG stops before $\alpha_0$ exceeds other couplings. At such
energies antilocalization regime does not occur.

\bibitem{Gonzalez}
J.~Gonzalez, F.~Guinea, and M.A.H.~Vozmediano, Nucl.\ Phys.\ B \textbf{424}, 596
(1994).

\bibitem{Foster08}
M.S.~Foster and I.L.~Aleiner, Phys.\ Rev.\ B \textbf{77}, 195413 (2008).

\bibitem{Foster06}
M.S.~Foster and A.W.W.~Ludwig, Phys.\ Rev.\ B \textbf{73}, 155104 (2006).

\bibitem{Peskin}
M.E.~Peskin and D.V.~Schroeder, \textit{An Introduction to Quantum Field
Theory,} Addison-Wesley Advanced Book Program (Westview Press, Boulder, 1995).

\bibitem{Bondi}
A.~Bondi, G.~Curci, G.~Pattuti, and P.~Rossi, Ann.\ Phys.\ \textbf{199}, 268
(1990).

\bibitem{Kivel}  A.N.~Vasil'ev, M.I.~Vyazovskii, S.E.~Derkachev, and
N.A.~Kivel', Theor.\ Math.\ Phys.\ \textbf{107}, 441 (1996).

\end{thebibliography}
\end{document}